\documentclass[11pt]{article}
\usepackage{mathtools}
\usepackage{graphicx}
\usepackage{url}
\usepackage{blkarray}
\usepackage{xcolor}
\usepackage{pbox}
\usepackage{amsmath,bm}
\usepackage{amsfonts}
\usepackage{float}
\usepackage{subfig}
\usepackage{titling}
\usepackage[font=small,labelfont =bf]{caption}

\begin{document}
\title{\bf An Introduction to Topological Data Analysis for Physicists: From LGM to FRBs}

\author{{\bf Jeff Murugan \& Duncan Robertson}
\\[0.5cm]{\it Laboratory for Quantum Gravity \& Strings}
\\{\it Department of Mathematics and Applied Mathematics}
 \\{\it University of Cape Town}}
\begin{titlingpage}

\maketitle
\begin{abstract}
\noindent Topological Data Analysis (TDA) is a novel, and relatively new approach to analysing 
high-dimensional data sets. It does this by focussing on global 
properties like the shape and connectivity of the data giving it a significant advantage over more conventional 
tools based on cluster analysis, a localised property of the data. However, some of its mathematical foundations, 
like algebraic topology and discrete Morse theory, are perceived as an intimidatingly steep upramp into the subject. 
Consequently, it has enjoyed much less popularity as a data-analysis tool than less abstract methods.
This article aims to change this. By focusing on a small set of simple examples, chosen primarily for their 
pedagogical value, we introduce and explain TDA's two principle branches; persistent
homology and the Mapper algorithm. We then illustrate the universality of the method by discussing its 
application to 
the intriguing data set of fast radio burst (FRB) observations. We close the article with a discussion of the 
resilience of topological data analysis to noise and some statistical and computational challenges faced by 
the method.
\end{abstract}
\end{titlingpage}

\tableofcontents
\newpage
\section{Introduction \& Motivation}
The late twentieth century saw the dawn of the age of the computer and with it, an unlocking of knowledge on a 
scale considered unimaginable even a few decades ago. All of this digital computing is predicated on Boolean 
algebra, the centuries old mathematics of binary logic. Today, we stand on the precipice of a new era, one that 
is dominated by data. From the Large Hadron Collider, to the Blue Brain Project, to the Square Kilometer 
Array, humans are generating data in unprecedented volumes. Indeed, it is hardly possible to read the news or 
turn on a computer without encountering reminders of the ubiquity of ``big data” sets in our modern world, and 
their myriad important implications for our lives and society at large. However, for the most part this data 
is complicated, noisy and cumbersome, and understanding how to extract the valuable information encoded 
therin remains the great challenge of data science.\\ 

Fortunately, this surge in data generation is commensurate 
with a corresponding improvement in computational power that has seen computers develop from the
room-occupying behemoths of the 1950's to the brink of quantum computing today. As a corollary however, this
rapid improvement in computational power also saw the emergence of increasingly higher dimensional data sets, 
where the number of explanatory variables for a particular data point far exceeds the sample data size
\cite{johnstone_statistical_2009}. Worse, modern data sets, like genetic sequencing data 
\cite{carlsson_topology_2009} or the temperature anisotropies in the cosmic microwave background, 
are typically very noisy and often plagued with missing values, making 
a traditional statistical analysis very difficult. \\
 
A common-sense starting point then, in dealing with such high-dimensional data sets is to reduce the
number of dimensions. The simplest method, {\it feature selection}, involves choosing
a subset of the explanatory variables, usually based on some domain-specific
knowledge. More versatile are {\it feature extraction} methods, which project the
dataspace into a lower dimensional subspace in such a way that most of the useful
information is preserved \cite{khalid_survey_2014}. The contraction of the
dataspace has another cost besides potential information loss: both linear and
non-linear feature extraction can lead to statistical models which are difficult
to interpret \cite{fodor_survey_2002}.

\subsection{Clustering algorithms}

In most applications, it is imperative to identify patterns in the data which
are informative in some way, after any initial dimensional reduction has
occurred. And while human beings are genetically hardwired to seek out patterns, manual pattern identification 
in large data sets is often impractical, hence the field of
{\it cluster analysis} emerged to automate this process \cite{jain_data_1999}.
A core assumption of clustering methods is that the distribution of points in dataspace is not uniform;
there exist clusters of points which share some measure of similarity. The task of the cluster algorithm then
is to identify such clusters, if they exist.\\

Most clustering algorithms take as input a dataset and a measure of dissimilarity between data points. The immediate output is the
assignment of one or more cluster identities to each observation. Ideally, these
clusters can then be used to construct a \textit{data abstraction}: a
representation of the core, relevant structure of the data in a humanly
comprehensible form. In practice however, this is difficult to achieve consistently
\cite{jain_data_1999}. Noisy, high-dimensional data poses particularly severe challenges to 
clustering analysis. It has proven to be difficult to create clustering algorithms which
are robust to noise and insensitive to initial paramater and metric choices
\cite{carlsson_topology_2009}.\\

Of particular relevance to us will be so-called {\it neighbour-based} clustering algorithms. 
These methods hinge on the construction of $\epsilon$-neighbouring
graphs, the nodes of which are observations. Undirected edges $\{i,j\}$ are
formed between a pair of observations $\bm{x}_{i}, \bm{x}_{j}$ if and only if the distance between them,
$\rho(\bm{x}_{i} - \bm{x}_{j}) < \epsilon$, for some real $\epsilon > 0$ and
metric $\rho$. Clusters are then identified as the disjoint, internally
connected subgraphs, which are separated from each other by some distance
greater than the $\epsilon$ threshold  \cite{hartmann_practical_2016}.

\begin{figure}[ht!]
  \centering \includegraphics[width=7cm]{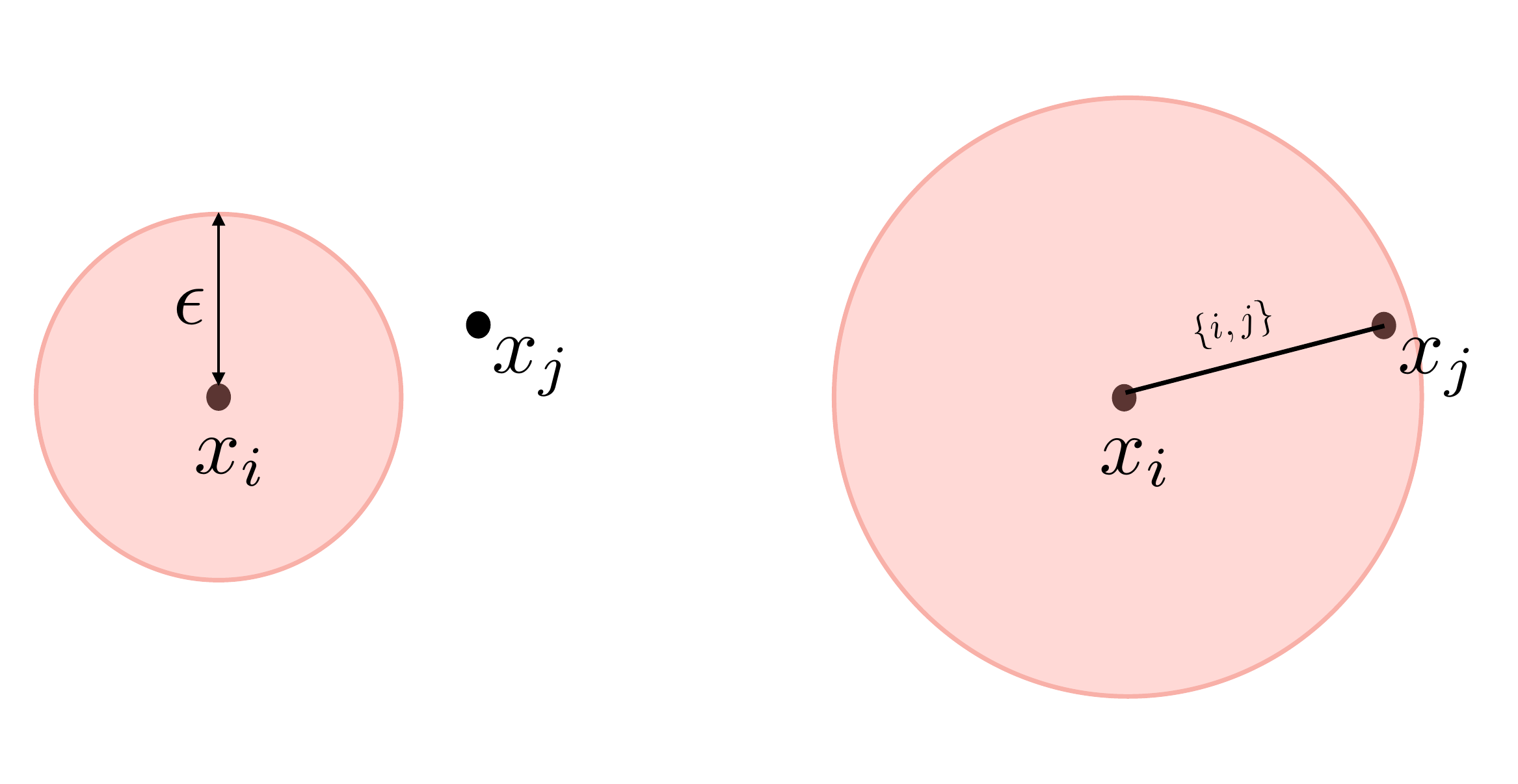}
  \caption{The formation of an edge between vertices as a function of $\epsilon$.}
  \label{edge}
\end{figure}

If $\epsilon$ is fixed, then the clustering method as described is
\textit{partitional}, in that the observations are partitioned into disjoint
clusters \cite{jain_data_1999}. For an $n$ node data set then, by varying $\epsilon$ 
continuously, from zero to some upper threshold value, we can observe how the cluster structure changes
from $n$ clusters of one point each, to a single one consisting of $n$ points. From this, we can form a {\it tree}
whose root is the single cluster and leaves, the individual observations. This construct is known as
a \textit{dendrogram}, a data abstraction constructed by a hierarchical
clustering procedure, known as single-linkage clustering
\cite{carlsson_topology_2009}.

\begin{figure}[ht!]
  \centering \includegraphics[width=10cm]{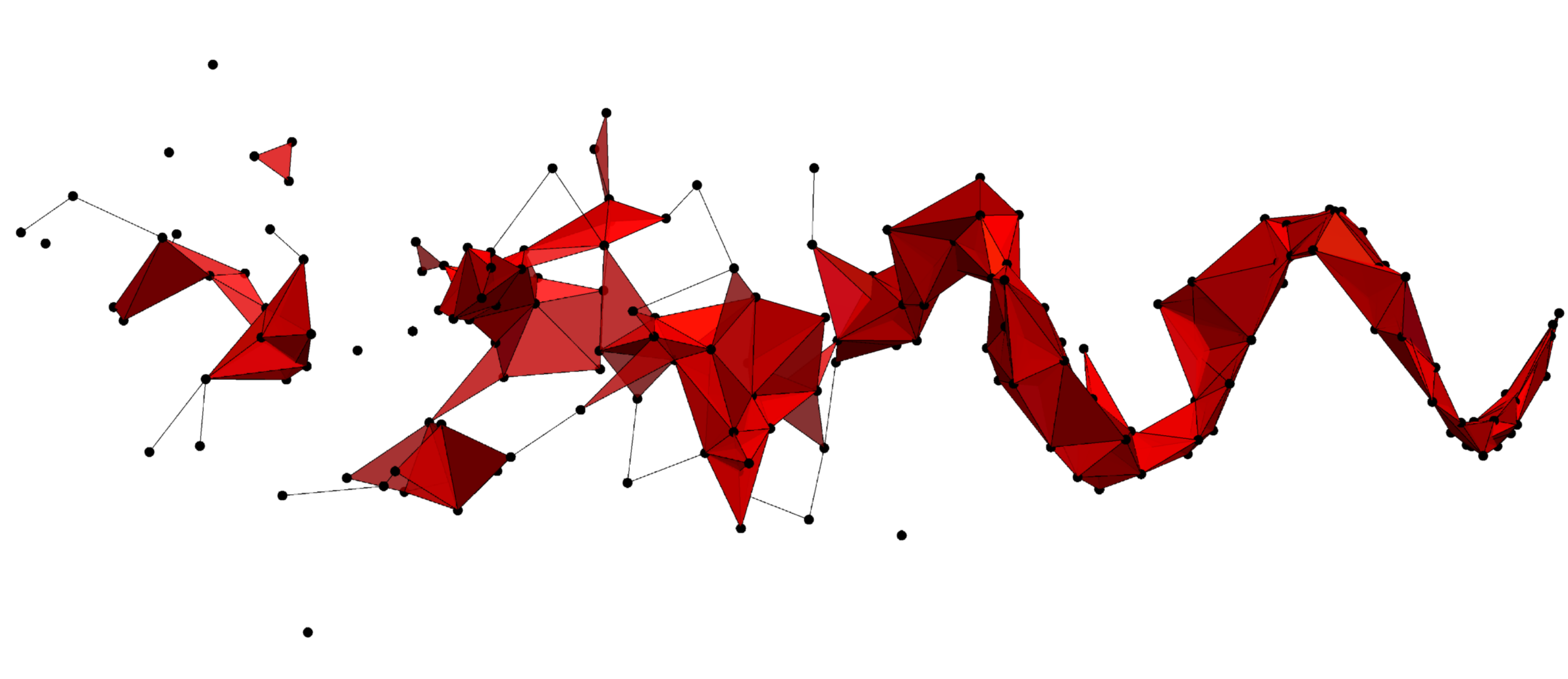}
  \caption{Data clustering and the formation of simplicies.}
\end{figure}

\subsection{Enter Topological Data Analysis}

The field of Topological Data Analysis (TDA) emerged in response to the various problems
in high-dimensional data analysis unresolved by clustering analysis. Topology focuses on
the \textit{qualitative} properties of geometric objects; as such, it is not as
sensitive to a choice of coordinates or choice of metric as purely geometric methods are \cite{carlsson_topology_2009}. This is famously, and glibly summed up in the slogan that topology does not distinguish between a coffee cup and the surface of a donut. Consequently, given some data set, the global strategy of TDA is to construct a geometric representation of the data and apply topological methods to that representation, from which inferences and data abstractions can be made about the shape and connectivity of the data \cite{chazal_introduction_2017}. The rest of this article will be devoted to making this notion precise and then applying it to some simple, illustrative examples that we have selected more for their pedagogical value than their originality, or relevance to the real world for that matter. In section 4, we deviate a little from this path by applying TDA to a particularly interesting data set in astronomy, the catalog of known observations of fast radio bursts (FRBs). With so little known about the physics of these astrophysical signals, we are particularly enamored by this, as far as we are aware, original application of TDA. Then, having extolled the many virtues of TDA, we conclude with a discussion of some of its statistical and computational challenges and an outlook to the (near) future.\\

Before we go on; a word on the philosophy of this paper. The elements of differential topology that constitute the first part of the text is well known to most theoretical physicists. However its application to a manifold of data points probably less so. Conversely, much of the language of data organisation in the latter part of the article is likely second nature to the data science community but mostly foreign to high energy theorists. To each of these communities, we apologise in advance for dwelling on what must seem like frustratingly trivial points. However, since our target audience is the, non-empty and rapidly growing, intersection of these communities, we found it necessary to err on the side of pedagogy. 

\section{Persistent Homology}

At this point, it is worth noting that there are two main facets to TDA: {\it persistent homology}, which is 
concerned with the classification and analysis of topological invariants associated to the data set, and 
{\it Mapper}, a powerful algorithm for creating direct visualisations of high-dimensional data
\cite{carlsson_topology_2009}. We will focus on each of these in turn in this and the following section.

\subsection{From data to a simplicial complex}

For our purposes, a data set $M$ will be taken to be a finite set of observation vectors $\bm{x}_{1},\bm{x}_{2}, \ldots \bm{x}_{n}$, each of length $d \in \mathbb{Z}_+$. A {\it metric} on this set is a function $\rho: M \times M \rightarrow \mathbb{R}_+$. We call the pair $(M, \rho)$ a (finite) metric space. While it is often convenient to view it as a finite sample from a continuous metric space \cite{chazal_introduction_2017}, we will not need to do so. For convenience, we will use the Euclidean metric for simple examples in this report. \\

Given the steep ramp in abstraction to follow, it will be worthwhile to proceed as intuitively as possible. Anyone who's watched a child colouring in a picture (as in Fig.3 for example) formed from some set of numbered dots observes (on average) a general strategy: join neighbouring dots to form recognisable shapes, then colour in various shapes bounded by the lines. Given a set of data points in some metric space, our task will be a generalisation of this: we will start by connecting the neighbouring points to form edges that bound certain shapes followed by `colouring in' the shapes.  In this context, the mode of colouring results in the creation of geometric objects called {\it simplices}.\\

\begin{figure}[ht!]
  \centering \includegraphics[width=6cm]{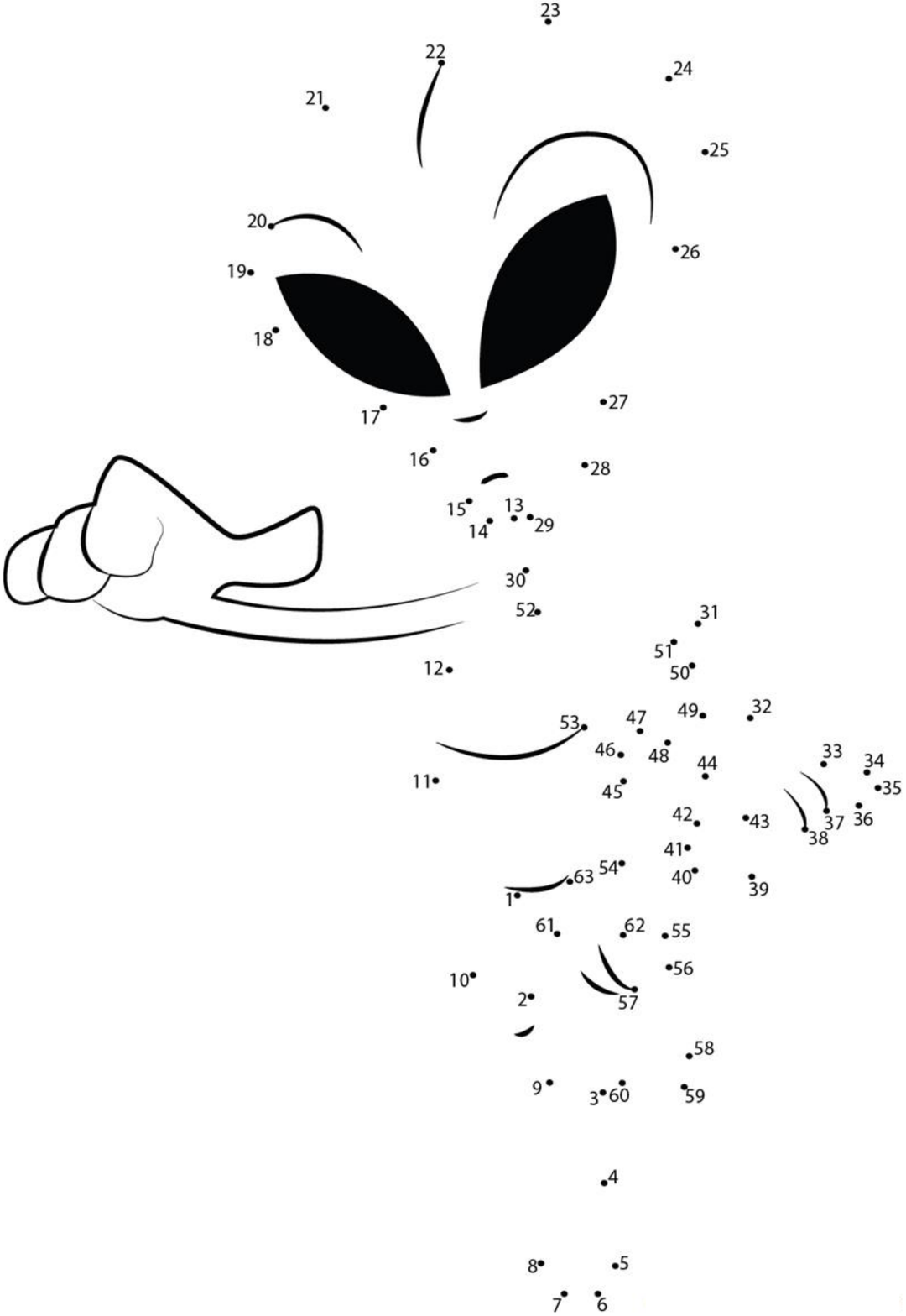}
  \caption{An example of a dot-to-dot colour-in puzzle (from http://www.connectthedots101.com)}
  \label{dot-to-dot}
\end{figure}
\noindent
A $k$-simplex is a set of $k + 1$ vertices which, we will denote $[ p_{0,}p_{1},\ldots, p_{k}]$. For example, the 0-simplex $[ p_{0}]$ is a point (or vertex), the 1-simplex $[ p_{1},p_{2}]$ is a line (or edge), the 2-simplex $[ p_{1},p_{2}, p_{3}]$ is a triangular disc ({\it i.e.} a triangle together with its enclosed area) and a 3-simplex $[ p_{1},p_{2}, p_{3},p_{4}]$, a regular tetrahedron. The \textit{faces} of a $k$-simplex $\sigma_{k}$ are all the simplices which are proper subsets of $\sigma_{k}$. For example, the faces of the tetrahedron in Fig. 4 include triangles, edges, vertices and the empty set, which is a face of every simplex.

\begin{figure}[ht!]
  \centering \includegraphics[width=5cm]{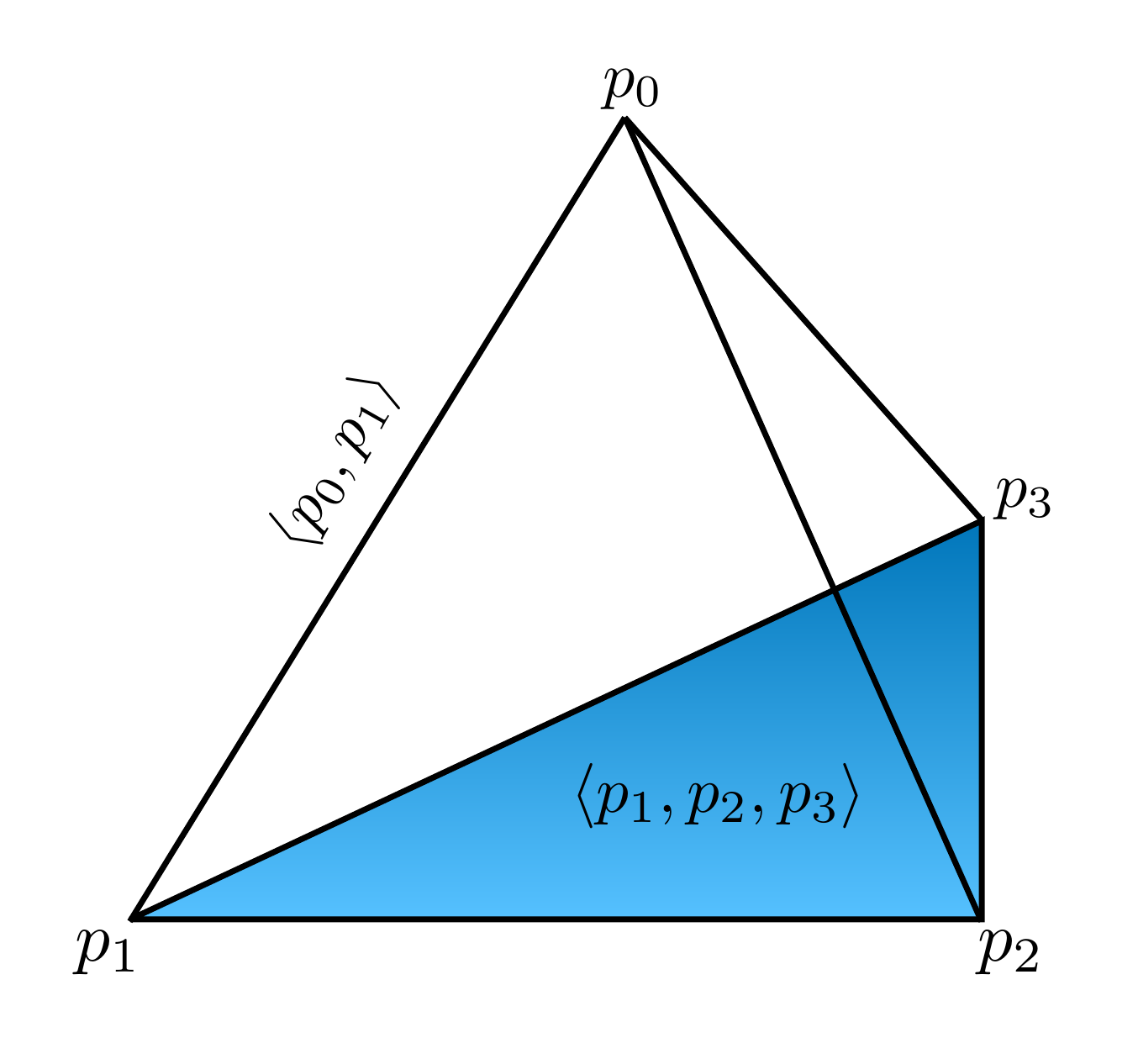}
  \caption{A tetrahedron as a collection of simplices. This particular collection contains one 3-face, four 2-faces, six 1-faces, four 0-faces and, of course, one open set. In general, there are $\binom{r+1}{p+1}$ $p$-faces in an $r$-simplex.}
  \label{tetra}
\end{figure}

\noindent
It will sometimes be convenient to have the simplex oriented. An oriented 1-simplex $\sigma_{1}$ is a {\it directed line segment} starting at $p_{0}$ and ending at $p_{1}$.  To disambiguate an oriented simplex from a disoriented one, we will denote the former by $\sigma_{1} = \langle p_{0},p_{1}\rangle$. In this notation then, an oriented 2-simplex $\sigma_{2} = \langle p_{0},p_{1},p_{2}\rangle$ is a triangular disk with a prescribed orientation along its boundary. To be concrete, we will adopt the convention that traversing the edges in a counter-clockwise sense counts as positive. As a result of our definition, if we denote a permutation of 0,1,2 by
\begin{eqnarray}
   P = \left(
      \begin{array}{ccc}
         0 & 1 & 2\\
         i & j & k
      \end{array}
   \right)\,,
\end{eqnarray}
then $\langle p_{i},p_{j},p_{k}\rangle = \mathrm{sgn}(P)\langle p_{0},p_{1},p_{2}\rangle$ where $\mathrm{sgn}(P) = \pm 1$ depending on whether $i,j,k$ is an even or odd permutation of 0,1,2. More generally, an oriented $r$-simplex,
\begin{eqnarray}
  \sigma_{r} = \langle p_{i_{1}},p_{i_{2}},\ldots,p_{i_{r}}\rangle = \mathrm{sgn}(P)\langle p_{1},p_{2},\ldots,p_{r}\rangle\,.
\end{eqnarray}
Our  goal will be to approximate the shape of some data set in an ambient data space. In order to do this, we need to be able to link simplices together in some systematic way to form `bigger' simplices. This process results in the construction of a \textit{geometric simplicial complex}, $\mathcal{K}$,  which is a collection of simplices satisfying the following two conditions \cite{chazal_introduction_2017}:
\begin{enumerate}
   \item $\mathcal{K}$ is closed under restriction in the sense that if $\tau$ is a face of $\sigma$ and 
   $\sigma \in \mathcal{K}$, then $\tau \in \mathcal{K}$.
   \item If $\sigma, \tau \in \mathcal{K}$ then $\sigma \cap \tau$ is either empty or a common face to both 
        $\sigma$ and $\tau$.
\end{enumerate} 
A $k$-complex is a geometric simplicial complex containing at least one $k$-simplex, and with no simplices of dimension strictly greater than $k$. For example, a {\it graph} is a 0-complex if it has no edges, and a 1-complex otherwise. In this sense, a $k$-complex can be considered a generalisation of a graph \cite{chazal_introduction_2017}. To construct a simplicial complex from data then, it is intuitive to use a generalisation of an $\epsilon$-neighbouring graph. The result is called a \textit{Vietoris-Rips complex} $\mathcal{V}_\epsilon(M)$;
\begin{equation*}
  \mathcal{V}_\epsilon(M) = \{\sigma \subseteq S | \rho(u,v) \leq \epsilon, \forall u \neq v \in \sigma\}\,,
\end{equation*}
\noindent where $\rho$ is the Euclidean metric. This simply means that we
construct a $k$-simplex from every collection of $(k+1)$ points which are pairwise
less than $\epsilon$ away from eachother \cite{zomorodian_fast_2010}. By this definition, it
follows that $\mathcal{V}_\epsilon(M)$ is not necessarily embedded in
$\mathbb{R}^d$, since the collection of $k+1$ points could well satisfy $k + 1 >
d$.

Once the simplicial complex is constructed, we can ask questions about its
topology. In particular, we can identify the presence of topological invariants
such as connected pieces, holes and cavities (2-dimensional holes). To do so, we will
need some tools from algebraic topology.

\subsection{From simplicial complex to homology groups}

\begin{figure}[ht!]
  \centering \includegraphics[width=\textwidth*1/2]{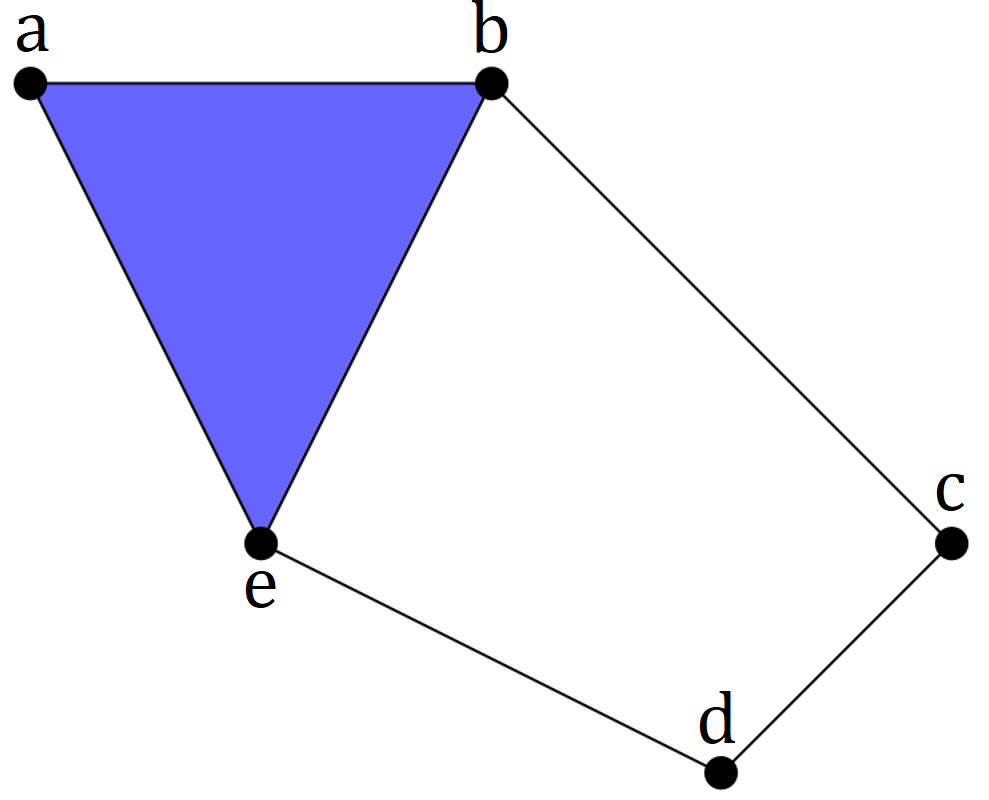}
  \caption{A simple 2-complex.}
  \label{complex}
\end{figure}

We will begin with a somewhat trivial example to demonstrate the construction of homology groups.
These in turn will be used to identify topological invariants, through the calculation
of \textit{Betti numbers}. Toward this end then, consider the simple 2-complex, $X$ in Fig.\ref{complex}.

We can write the complex above as the set $X = \{a, b, c, d, e,\langle a, b
\rangle,\langle b, c \rangle,\langle c,d \rangle$, $\langle d, e \rangle,\langle e, a
\rangle,\langle b, e \rangle, \langle a, b, e \rangle\}$, where as usual, $\langle a_1,
a_2, ...,a_k \rangle = \sigma$ denotes an (unoriented) $k$-simplex.

\subsubsection{Chain group}
In what follows, it will also prove useful to define the operation of addition of simplices, which can be thought of as the result of  `gluing' simplices together to form a sub-complex. For example, $\langle a,b \rangle + \langle b,c \rangle$ is a 1-complex in $X$ formed by those edges. Notice that their intersection, $\langle a,b \rangle \cap \langle b,c \rangle = b \subset X$. Using this operation, we can define a $k$-chain of a simplicial complex $S$ as the formal sum of $k$-simplices, i.e. if $c$ is $k$-chain, then
\begin{equation*}
  c = \sum a_i \sigma_i,
\end{equation*}
where $\sigma_i$ is a $k$-simplex and  the coefficients $a_i $ take values in some field, $F$, which we will usually chose to be the cyclic group or order two, $\mathbb{Z}_2$, for computational simplicity \cite{edelsbrunner_computational_2010}. We will denote the set of all $k$-chains of a simplicial complex $S$ by $C_k(S)$. We can then form a group by defining addition of chains component-wise, i.e. if $c = \sum c_i \sigma_i$ and $d = \sum d_i\sigma_i$ are $k$-chains, then so is $c + d = \sum (c_i + d_i) \sigma_i$. It is easy to see that $C_k(S)$ forms a group, called the \textit{$k^{\text{th}}$ chain group}, under this operation. By convention, we will use $C_k(S)$ as a shorthand for the group $(C_k(S), +)$.  \\

$C_k(S)$ is an abelian group because its group operation is based on commutative, component-wise addition. 
        In fact, $C_k(S)$ is the \textit{free abelian group} generated by the set of k-cycles in S, a fact which follows directly from the definition of a chain \cite{munkres_elements_1984}. Thus, if the set of $k$-cycles is $\{\sigma_1, \sigma_2, \dots \sigma_n \}$, then
\begin{equation*}
   C_k(S) = \text{span}(\sigma_1, \sigma_2, \dots \sigma_n).  
\end{equation*}
 Hence, the rank of $C_k(S)$, i.e. the number of its basis elements, is $n$. For example, for $X$ as above and with $F = \mathbb{Z}_2$, we have only one 2-simplex ($\langle a,b,e \rangle$) and hence
\begin{equation*}
  C_2(X) = \text{span}(\langle a,b,e \rangle)= \{\lambda\langle a,b,e \rangle \ |\  \lambda \in \mathbb{Z}_2\} = \{\langle a,b,e \rangle, 0\}\simeq \mathbb{Z}_{2}.
\end{equation*}
Clearly, $\text{rank}\left(C_2(X)\right) = 1$. Intuitively, chains should be thought of as the natural objects that pair with integrands under the (definite) integral sign to produce a number.   

\subsubsection{Boundary operator}
The next tool that we will need is a group homomorphism called \textit{the boundary operator}:
\begin{equation*}
\partial_{k}: C_k(S) \rightarrow C_{k-1}(S) \text{, for } k \geq 1, 
\end{equation*}
Intuitively, the
boundary of a $k$-simplex is the set of all of its $(k-1)$-dimensional faces.
Hence, the boundary of a 2-simplex is its perimeter edges and the boundary of an
edge is the difference of its vertices. Formally, we define the boundary of a
$k$-simplex $\sigma_{k}$ with vertices $\langle v_0, v_1, ... v_k \rangle$ as
\begin{equation*}
  \partial_{k}(\sigma_{k}) = \sum_{i=0}^k (-1)^i \langle v_0, ..., \widehat{v}_i, ..., v_k \rangle,
\end{equation*}
\noindent where $\widehat{v}_i$ is a vertex removed from $\sigma_{k}$. For our $X$ example,
\begin{equation*}
\partial_{2}(\langle a,b,e \rangle) = \langle b,e \rangle - \langle a,e \rangle + \langle a,b \rangle = \langle b,e \rangle + \langle e,a \rangle + \langle a,b \rangle.
\end{equation*}
The boundary map has two important properties worth mentioning:
\begin{enumerate}
\item The boundary of a vertex is zero: $\partial_{0}(v_i) = 0 \quad \forall v_i \in
  \sigma$.
\item The boundary operator is {\it nilpotent} in the sense that the boundary of a boundary is zero: $\partial^{2}(\sigma) \equiv \partial(\partial(\sigma)) = 0 \quad\forall \sigma \in S$.
\end{enumerate}
These properties allow us to identify {\it k-cycles} in a simplicial complex $S$ as those $k$-simplices $\sigma_{k} \in S$ without boundary {\it i.e.} for which $\partial(\sigma_{k}) = 0$. For each chain $C_k(S)$, we define $Z_k$ to be the set of all $k$-cycles. In other words, $Z_k$ is the kernel of the map $\partial_{k}: C_k(S) \rightarrow C_{k-1}(S)$. It follows that $Z_k$ forms a subgroup of $C_k(S)$ \cite{munkres_elements_1984}.\\

\noindent
Any cycle that can be expressed as the boundary of a higher dimensional cycle we will call a {\it boundary}, and denote the set of all $k$-boundaries by $B_{k}$. It is easily shown that this set forms a subgroup of $C_k(S)$ under the usual group addition. In particular, since $\partial^{2}(\sigma) = 0$ for all $\sigma \in C_{k+1}(S)$, it follows that $B_{k} = \text{Im}(\partial(C_{k+1}))$ is a subgroup of the kernel of $\partial: C_k(S) \rightarrow C_{k-1}(S)$, i.e. $B_{k} \leq Z_k $. In words, this is the statement that {\it every boundary is also a cycle}. This is trivially true since the boundary operator is nilpotent. The converse question is more interesting; when is a cycle a boundary? 

\subsubsection{Homology groups and Betti numbers}
We have seen that, given some $n$-complex $S$ there exists nested subgroups $B_k \leq Z_k \leq C_k$ for each  $k = 0 \ldots n$. Since $C_{k}$ is abelian, all the subgroups are {\it normal}. A $k$-dimensional hole in $S$ is a $k$-cycle that is not a boundary of a $k+1$-dimensional simplex. So, if we want to enumerate the number of $k$-holes in $S$, we need to identify the $k$-cycles in $Z_k(S)$ that are not the boundary of a higher dimensional simplex. Hence, we want the group of cycles modulo boundaries. This goes by the name of the $k^{th}$ {\it homology group},
\begin{equation*}
  H_k(S) = Z_k(S) / B_k(S) = \text{Ker}(\partial_{k}(C_k)) / \text{Im}(\partial_{k+1}(C_{k+1}))\,.
\end{equation*}
The elements of $H_k(S)$ are cosets of the form $\beta + B_k,$ for $\beta\in Z_k$. These are called \textit{homology classes}, which form a group under the operation
\begin{equation*}
   (B_k + \alpha) + (B_k + \beta) = B_k + (\alpha + \beta)\,,
\end{equation*}
for all $\alpha,\beta \in Z_k$. The rank of $H_k(S)$ is called the \textit{$k^{th}$ Betti number} and denoted $b_k$. In other words, $b_k$ is the number of homology classes of $Z_k$, {\it i.e.} the number of $k$-dimensional holes in $S$. The zero'th Betti number, $b_{0}$, in particular gives the number of connected components of $S$. It is a useful fact of life that $\text{rank}(H_k) = \text{rank}(Z_k) - \text{rank}(B_k)$ \cite{edelsbrunner_computational_2010}. To illustrate some of these abstract concepts, let's do some examples.

\begin{enumerate}
\item {\bf An intuitive example - the torus:}
Before embarking on a systematic computation of the Betti numbers for a simplicial complex, let's first develop some intuition for the problem with a familiar example, the torus $T^{2}$ (see Fig.6).

\begin{figure}[ht!]
  \centering \includegraphics[width=\textwidth*1/2]{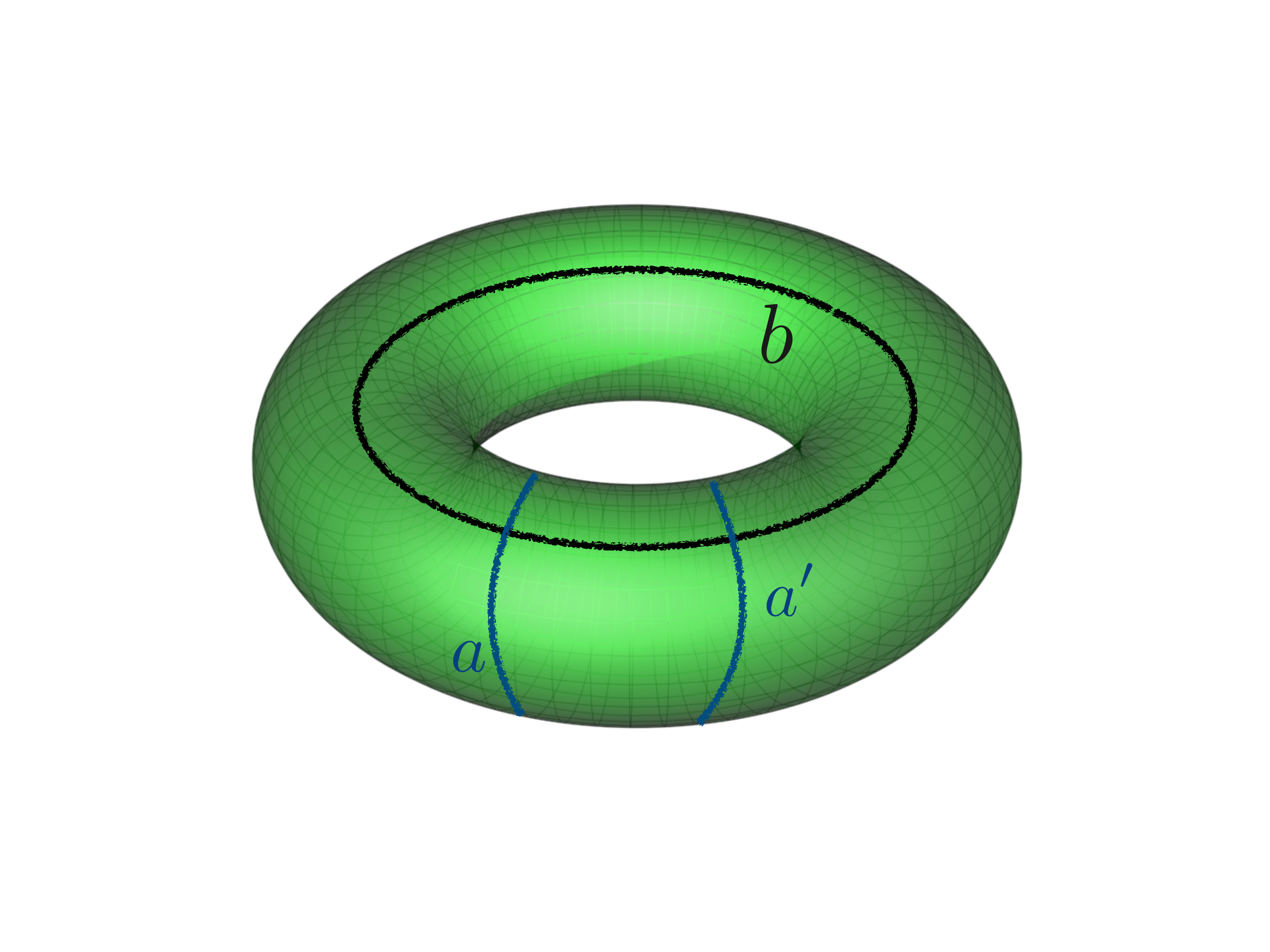}
  \caption{A green 2-Torus with some 1-cycles.}
  \label{torus}
\end{figure}

\noindent
Recall that the $k$'th homology group is generated by  boundaryless $k$-chains that are not themselves the boundaries of some $(k+1)$-chains, with the $k$th Betti number counting the number of such generators. The torus is a 2-dimensional surface without a boundary ({\it i.e.} $\partial_{2}T^{2} = 0$), but is clearly not the boundary of any 3-chain\footnote{Contrary to popular analogy, the 2-torus is of course the {\it surface} of a donut and not a solid donut.}. As a result, $H_{2}(T^{2})$ is generated by just one generator, the torus itself and we identify $H_{2}(T^{2})\cong \mathbb{Z}$. Since the torus is a connected surface, it has just one connected component and $H_{0}(T^{2})\cong \mathbb{Z}$. To compute $H_{1}$, let's consider the loops depicted in Fig.6. Clearly these closed loops have no boundaries but are not themselves boundaries of any 2-chain. Now consider the two loops $a$ and $a'$. Since $a' - a$ is the boudary of the cylindrical region inbetween, $a'$ is {\it homologous} to $a$ {\it i.e.} they belong to the same homology class. Similar arguments hold for the 1-cycle $b$. Evidently then, $H_{1}(T^{2})$ is freely generated by the independent 1-cycles $a$ and $b$, and $H_{1}(T^{2})\cong \mathbb{Z}\oplus \mathbb{Z}$. Given the homology groups, by counting the number of generators we can immediately read off that $b_{0}(T^{2}) = 1, b_{1}(T^{2}) = 2, b_{2}(T^{2}) = 1$.
  
\item{\bf Another X-ample - finding Betti numbers laboriously:}
Let's return to our example of the 2-complex $X$ in Fig.5. using the field $\mathbb{Z}$ for now. It is easiest to compute the highest order Betti number first, then work our way down to $b_0$. Recall that the group of 2-chains in $X$, $C_2(X) = \alpha_0\langle a,b,e \rangle$ for $\alpha_0 \in\mathbb{Z}$. Linearity of the boundary operator means that,
\begin{align*}
  \partial_{2}(\alpha_0 \langle a,b,e \rangle) &= \alpha_0 \cdot \partial_{2}(\langle a,b,e \rangle) \\
                             &= \alpha_0 (\langle b,e \rangle + \langle e,a \rangle + \langle a,b \rangle) \\
                             &= 0 \iff \alpha_0 = 0.
\end{align*}                            
In other words, Ker$(\partial_{2}(C_2(X)) = \{0\} = Z_2$. Since there are no higher dimensional simplices, $C_3(X) = \emptyset$, hence Im $\partial_{3}(C_3(X)) = \{0\} = B_2$. Consequently, the quotient group $H_2 = Z_{2}/B_{2} = \{0\}$. Hence, $b_2 = \text{rank}(H_2) = 0$. This is the statement that there are {\it no cavities (2-dimensional holes) in X}, as we would expect by looking at Fig.5.\\

\noindent
Moving one dimension down, calculating the kernel of $\partial_{1}$ acting on $C_1(X)$ is a messier affair. Nevertheless, we persist. If $\lambda_i \in \mathbb{Z}\ \forall i$, then
\begin{eqnarray*}
  \partial_{1}(C_1(X)) &=& \lambda_0(a-b) + \lambda_1(b-c) + \lambda_2(c-d) +
  \lambda_3(d-e)\\
   &+& \lambda_4(e-a) + \lambda_5(b - e) \\
  &=& a(\lambda_0 - \lambda_4) + b(\lambda_1 - \lambda_0 + \lambda_5) +
  c(\lambda_2 - \lambda_1)\\
   &+& d(\lambda_3 - \lambda_2) + e(\lambda_4 - \lambda_3
  - \lambda_5) .
\end{eqnarray*}
To find the kernel of the map, we need to find the constraints on the $\lambda_{i}$ for which $\partial_{1}(C_{1})$ vanishes. It is not difficult to see, from the second line above that this is true when $\lambda_0 = \lambda_4$; $\lambda_1 = \lambda_2 = \lambda_3$ and $\lambda_5 = \lambda_0 - \lambda_1$. With a little rearrangement then, we can read off that
\begin{equation*}
  \text{Ker}(\partial_{1}(C_1)) = \lambda_0(\langle a,b \rangle + \langle b,e \rangle + \langle e,a \rangle) +
  \lambda_1(\langle b,c \rangle + \langle c,d \rangle + \langle d,e \rangle + \langle e,b \rangle) = Z_1\,. 
\end{equation*}
The 1-cycles of $X$ are the boundary of the triangle $\langle a,b,c \rangle$, and the
boundary of the rectangle with vertices $a,b,c$ and $d$, as expected. Since $Z_1
$ has two elements, it is isomorphic to $\mathbb{Z}_2$. Now we need to find
$B_1$:
\begin{alignat*}{2}
  & & B_1 = \text{Im}(\partial_{2}(C_2)) &= \beta_0(\langle b,e \rangle - \langle a,e \rangle + \langle a,b \rangle) \\
  & & &= \beta_0(\langle a,b \rangle + \langle b,e \rangle + \langle e,a \rangle),
\end{alignat*}
hence $B_1$ is the boundary of the triangle $\langle a,b,c \rangle$. Finally, we have that
\begin{align*}
  H_1 &= Z_1 / B_1 \\
      &= \text{span}(\langle b,c \rangle + \langle c,d \rangle + \langle d,e \rangle + \langle e,b \rangle ) \cong \mathbb{Z}.
\end{align*}
We interpret this as the statement that the only 1-cycle of X which is not the boundary of a higher order simplex
is the boundary of the rectangle. Again, this is intuitively clear from Fig.5. It
follows that $b_1 = \text{rank}(H_1) = 1$, {\it i.e.} there is one 1-dimensional hole
in X.\\

\noindent
Finally, we need to compute the number of connected components of X. Firstly, since 
\begin{equation*}
  \partial(C_0(X)) = 0  \Rightarrow \text{Ker}(\partial_{0}(C_0))) = Z_0 = C_0 = \text{span}(a,b,c,d,e)\,.  \cong \mathbb{Z}^5
\end{equation*}
Now, to compute $B_0$ notice that,
\begin{equation*}
  B_0 = \text{Im}(\partial_{1}(C_1)) = \text{span}(a-b, b-c, c-d, d-e, e-a, b-e)\,.
\end{equation*}
To clarify this a little, notice that since we can write $b - e = -\big((a - b) + (e - a)\big)$ and $c - d = -\big((d -
e) + (e - a) + (a - b) + (b - c)\big)$,  the combinations $b -e$ and $c - d$ in fact lie in the span
of the basis $\{a-b,b-c,e-a,d-e\}$, {\it i.e.} $B_0 \cong \mathbb{Z}^4$.
It follows then that $H_0 = Z_0/B_0 \cong \mathbb{Z}^5/\mathbb{Z}^4 = \mathbb{Z}$ so
that $b_0 = \text{rank}(\mathbb{Z}) = 1$, {\it i.e.} there is one connected component
of X.

\item {\bf A disorienting example - the M\"obius band:} As a final illustrative example, we'll now consider a non-orientable example; the M\"obius band. As every school child knows, this famously one-sided surface can be made by first twisting, and then gluing together two sides of a rectangle. This gives us a clue as to how to triangulate the surface to an associated simplicial complex, $\mathfrak{M}$, shown in Fig.7. From this we can now compute the Betti numbers as follows. Starting from the observation that, oriented or not, the M\"obius band is connected, we know immediately that $H_{0}(\mathfrak{M}) = \left\{\alpha[p_{i}]|\alpha\in\mathbb{Z}\right\}\cong \mathbb{Z}$, where $p_{i}$ is any 0-simplex of $\mathfrak{M}$.

\begin{figure}[ht!]
  \centering \includegraphics[width=\textwidth]{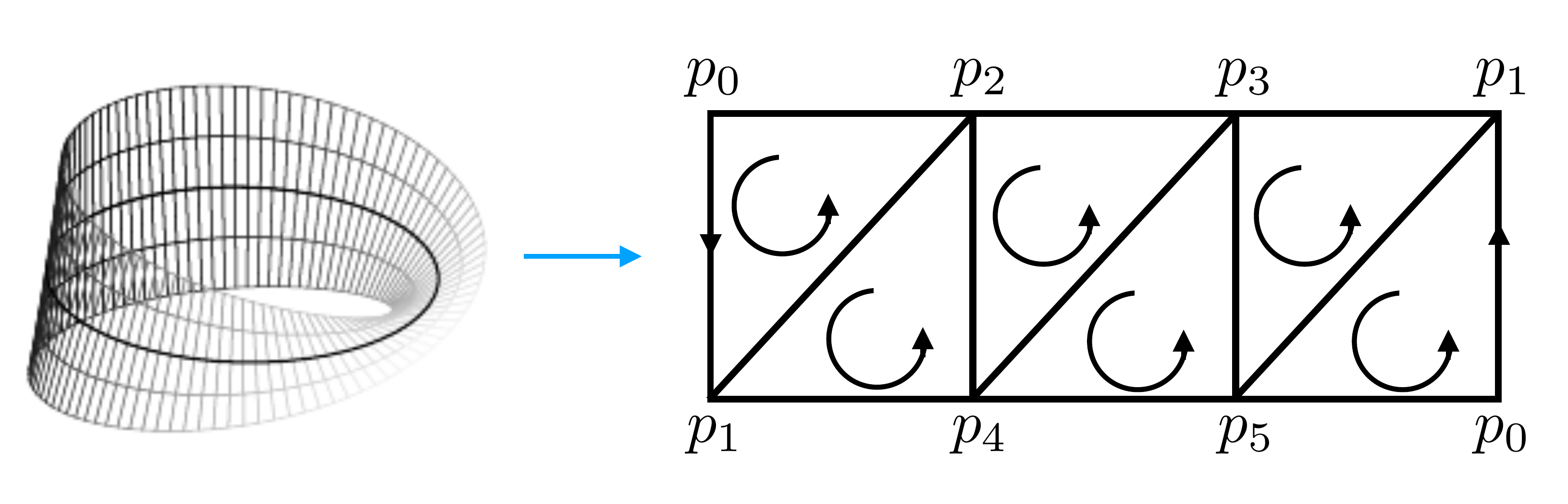}
  \caption{A M\"obius band and its associated simplicial complex.}
  \label{mobius}
\end{figure}

\noindent
Next, since the surface is 2-dimensional, $B_{2}(\mathfrak{M}) = \{0\}$. A 2-cycle $z\in Z_{2}(\mathfrak{M})$ can be writen in the form (with the orientation indicated in Fig.7)
\begin{eqnarray*}
   z &=& i\langle p_{0},p_{1},p_{2}\rangle +  j\langle p_{2},p_{1},p_{4}\rangle +  k\langle p_{2},p_{4},p_{3}\rangle \\
       &+&  l\langle p_{3},p_{4},p_{5}\rangle +  m\langle p_{3},p_{5},p_{1}\rangle 
       + n\langle p_{1},p_{5},p_{0}\rangle\,.
\end{eqnarray*}
Acting with the boundary operator,
\begin{eqnarray*}
   \partial_{2}z &=& i\left\{\langle p_{1},p_{2}\rangle - \langle p_{0},p_{2} + \langle p_{0},p_{1}\rangle\rangle\right\}\\
       &+& j\left\{\langle p_{1},p_{4}\rangle - \langle p_{2},p_{4} + \langle p_{2},p_{1}\rangle\rangle\right\}\\
       &+&k\left\{\langle p_{4},p_{3}\rangle - \langle p_{2},p_{3} + \langle p_{2},p_{4}\rangle\rangle\right\}\\
       &+&l\left\{\langle p_{4},p_{5}\rangle - \langle p_{3},p_{5} + \langle p_{3},p_{4}\rangle\rangle\right\}\\
       &+& m\left\{\langle p_{5},p_{1}\rangle - \langle p_{3},p_{1} + \langle p_{3},p_{5}\rangle\rangle\right\}\\
       &+&n\left\{\langle p_{5},p_{0}\rangle - \langle p_{1},p_{0} + \langle p_{1},p_{5}\rangle\rangle\right\} = 0\,,
\end{eqnarray*}
if, and only if all the coefficients $i =j =k =l =m =n =0$, establishing that $Z_{2}(\mathfrak{M}) = \{0\}$. Consequently, $H_{2}(\mathfrak{M}) = Z_{2}(\mathfrak{M}/B_{2}(\mathfrak{M}\cong \{0\}$ also. As you might suspect, the first homology group is the most tedious to compute by brute force. Since we're looking for loops that are not themselves the boundary of any 2-simplex, intuitively we would guess that there is only one class of such loops (see Fig.7). To confirm our guess, we can take as a representative 1-cycle,
\begin{eqnarray*}
   z = \langle p_{0},p_{1}\rangle + \langle p_{1},p_{4}\rangle + \langle p_{4},p_{5}\rangle 
   + \langle p_{5},p_{0}\rangle\,.
\end{eqnarray*}
Had we chosen, say
\begin{eqnarray*}
   \widetilde{z} = \langle p_{0},p_{2}\rangle + \langle p_{2},p_{3}\rangle + \langle p_{3},p_{5}\rangle 
   + \langle p_{5},p_{1}\rangle\,,
\end{eqnarray*}
then we would have found that 
\begin{eqnarray*}
   z - \widetilde{z} = \partial_{2}\left\{\langle p_{2},p_{1},p_{4}\rangle + \langle p_{3},p_{4},p_{5}\rangle + 
   \langle p_{1},p_{5},p_{0}\rangle + \langle p_{2},p_{4},p_{3}\rangle\right\}\,.
\end{eqnarray*}
In other words, $\widetilde{z}$ is homologous to $z$. In fact, with a little bit of effort it can be shown that, in fact all closed loops are homologous to $nz$ for some $n\in\mathbb{Z}$. Another way of saying this is that $H_{1}(\mathfrak{M})$ is generated by just one element, $[z]$, or $H_{1}(\mathfrak{M})\cong\mathbb{Z}$. Now, counting generators gives $b_{0} = 1, b_{1} = 1, b_{2} = 0$ for the M\"obius band.
\end{enumerate}

\subsubsection{The Euler-Poincar\'e formula}
The Euler characteristic is, arguably, the most familiar of topological invariants. Since this familiarity stems largely from our well-honed intuition for shapes in $\mathbb{R}^{3}$, we'll begin there, before generalising to higher dimensional objects in higher dimensional spaces. A {\it polyhedron} is a geometrical shape consisting of faces, which meet at faces. Faces, in turn, meet at vertices. Of course, if this sounds familiar, it is because a polyhedron is nothing but a special type of simplicial complex in $\mathbb{R}^{3}$. The {\it Euler characteristic} of any subset $X$ of $\mathbb{R}^{3}$ that is homeomorphic to a polyhedron $\mathfrak{P}$ is computed as
\begin{eqnarray*}
   \chi(X) \equiv V - E + F\,,
\end{eqnarray*} 
where $V$ counts the number of vertices, $E$ the number of edges and $F$ the number of faces in the polyhedron $\mathfrak{P}$. Some examples:
\begin{itemize}
   \item Since simplest polyhedron homeomorphic to a circle $S^{1}$ is a triangle, $\chi(\bigcirc) = \chi(\bigtriangleup) = 3 - 3 = 0$. Note that, had we chosen a different polyhedron, say a square, we would have found that $\chi(\Box) = 4-4 = 0$. This is consistent with the fact that the Euler is a topological invariant that is blind to local (geometric deformations) such as, in this case, blowing up one vertex into two vertices connected by an edge.
   \item Going up one dimension, the Euler characteristic of the 2-sphere can be computed by mapping $S^{2}$ to an associated polyhedron like a cube. Then $\chi(S^{2}) = \chi(\mathrm{cube}) = 8 - 12 + 6 = 2$. Again, this number stays the same whether we use the cube, or a tetrahedron or indeed {\it any} polyhedron homeomorphic to the 2-sphere. This is in fact guarenteed by the {\it Poincar\'e-Alexander theorem}.
   \item Slightly less trivial is the case of the 2-torus. An example of a polyhedron homeomorphic to $T^{2}$ is given in Fig.8.
   \begin{figure}[ht!]
      \centering \includegraphics[width=\textwidth*1/2]{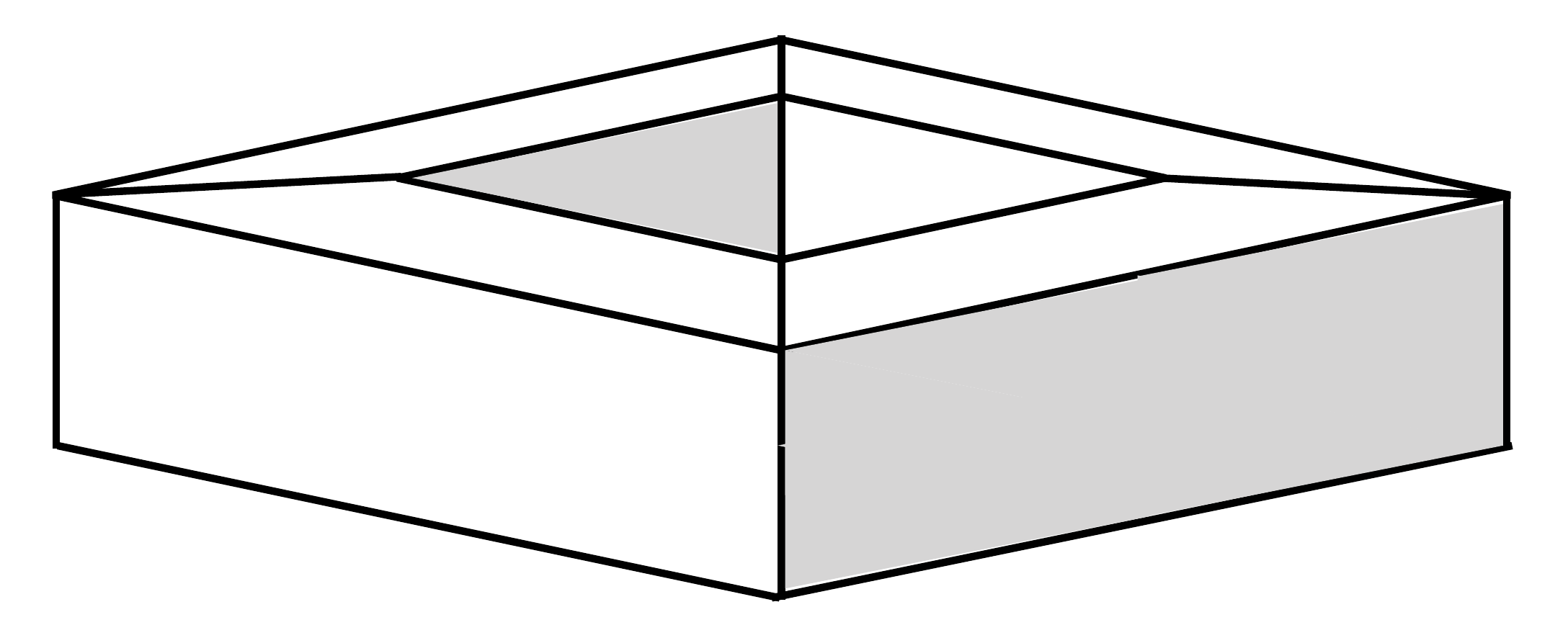}
      \caption{A polyhedron homeomorphic to the 2-torus.}
     \label{torus2}
   \end{figure}
   Counting its faces, edges and vertices gives $\chi(T^{2}) = 16 - 32 + 16 = 0$.
\end{itemize}
To generalize the Euler characteristic to complexes beyond polyhedra embedded in $\mathbb{R}^{3}$, as we will need to do soon enough, note that $\chi(\mathfrak{P})$ is the alternating sum over the number of simplices of all dimensions up to the dimension of $\mathfrak{P}$. More precisely, if $K$ is an $n$-dimensional simplicial complex and $I_{r}$ is the number of $r$-simplices in $K$,
\begin{eqnarray*}
   \chi(K) = \sum_{r=0}^{n}(-1)^{r}I_{r}\,.
\end{eqnarray*}
Since $\partial_{r}$ is a map between the vector spaces $C_{r}$ and $C_{r-1}$, $I_{r} = \mathrm{rank}(C_{r}) = \mathrm{rank}(\mathrm{Ker(\partial_{r})}) + \mathrm{rank}(\mathrm{Im(\partial_{r})}) = \mathrm{rank}(Z_{r}) + \mathrm{rank}(B_{r-1})$. On the other hand, we know that $\mathrm{rank}(H_{r}) = \mathrm{rank}(Z_{r}) - \mathrm{rank}(B_{r})$. These two equations, together with the fact that $\mathrm{rank}(B_{0}) = \mathrm{rank}(B_{n}) = 0$, allows us to write the Euler characteristic in terms of the Betti numbers as,
\begin{eqnarray*}
   \chi(K) = \sum_{r=0}^{n}(-1)^{r}b_{r}(K)\,.
\end{eqnarray*}
This remarkable result - the {\it Euler-Poincar\'e} formula - makes it clear that the Euler characteristic is a topological invariant, since it can be expressed directly as a sum over topological invariants . As a sanity check, using the Betti numbers from our earlier computation, we find that $\chi(T^{2}) = 1 - 2 + 1 = 0$, as expected.

\subsubsection{X revisited - computing Betti numbers with linear algebra}
\label{bettimatrix}
As fun as it is, clearly the manual computation of Betti numbers is an arduous task, that we expect to get even more so as we proceed to high-dimensional data sets. Fortunately, there is a more elegant alternative that will be particularly useful to implement algorithmically. Key to this is a matrix representation of the boundary operator $\partial_{k}\!:\! C_{k}(S)\to C_{k-1}(S)$. The columns of this matrix are spanned by the $k$-simplices of $S$, while the rows are spanned by the $(k-1)$-simplices of S. Using $\mathbb{Z}_2$ as the field for $C_k(S)$, if we let $\sigma_{k,j}$ be the $j^\text{th}$ simplex of order $k$, then the entries of $\partial_k = [d_{ij}]$ are such that
\begin{equation*}
    d_{ij} = \begin{cases} 
        1 & \text{if }\sigma_{k-1, i} \text{ is a face of } \sigma_{k, j}, \\
      0 & \text{otherwise}.
    \end{cases}
\end{equation*}
This means that, given a $k$-simplex written as a vector $\bm{c}$, the boundary of $\bm{c}$ is simply 
$\partial_k \bm{c}$. For our example complex $X$, we can write the matrix representation of the boundary map taking 1-simplices to 0-simplices is  
\begin{equation*}
  \partial_1 = 
  \begin{blockarray}{ccccccc}
     & \langle a,b \rangle & \langle b,c \rangle & \langle c, d \rangle & \langle d,e \rangle & \langle e,a \rangle & \langle b,e \rangle \\
    \begin{block}{c(cccccc)}
      a & 1 & 0 & 0 & 0 & 1 & 0 \\
      b & 1 & 1 & 0 & 0 & 0 & 1 \\
      c & 0 & 1 & 1 & 0 & 0 & 0 \\
      d & 0 & 0 & 1 & 1 & 0 & 0 \\
      e & 0 & 0 & 0 & 1 & 1 & 1 \\
    \end{block}
  \end{blockarray}\,.
\end{equation*}
With this, the boundary of the chain, say, $\langle a,b \rangle + \langle b,c \rangle$ is computed as
\begin{equation*}
\partial_1
\left(
   \begin{array}{c}
   1\\
   1\\
   0\\
   0\\
   0\\
   0
   \end{array}
\right) = 
\left( \begin{array}{c}
   1\\
   1+1\\
   1\\
   0\\
   0 
\end{array}
\right)
 = \left( \begin{array}{c}
   1\\
   0\\
   1\\
   0\\
   0 
\end{array}
\right) = 
a + c,
\end{equation*}
counting modulo 2. Already a significant computational improvement over the homology calculations of the previous section, this can be further simplified if the chain modules of the complex are {\it finitely generated}, for example, if the field over which the simplicial complex is defined is $\mathbb{Z}$ or $\mathbb{Z}_{2}$. In this case, the (rectangular) boundary matrix can be reduced to its \textit{Smith normal form} (SNF) consisting of a square diagonal submatrix in the upper left corner and zeroes elsewhere. This is obtained from the original matrix by multiplying on the left and right by square invertible matrices. Equivalently, the boundary matrix can be diagonalised by a sequence of row and column operations that leave its rank invariant. These operations are nothing but the {\it Gauss reduction} of a system of linear equations that every high school student is familiar with. For an 
       $n_{k-1}\times n_{k}$ boundary matrix, the reduction procedure requires a runtime that is at most cubic, and an amount of memory that is at most quadratic in the number of simplices in the simplicial complex. In other words, Gaussian elimination is remarkably efficient at reducing the boundary matrix to its SNF.\\

\begin{figure}[ht!]
  \centering \includegraphics[width=\textwidth*2/4]{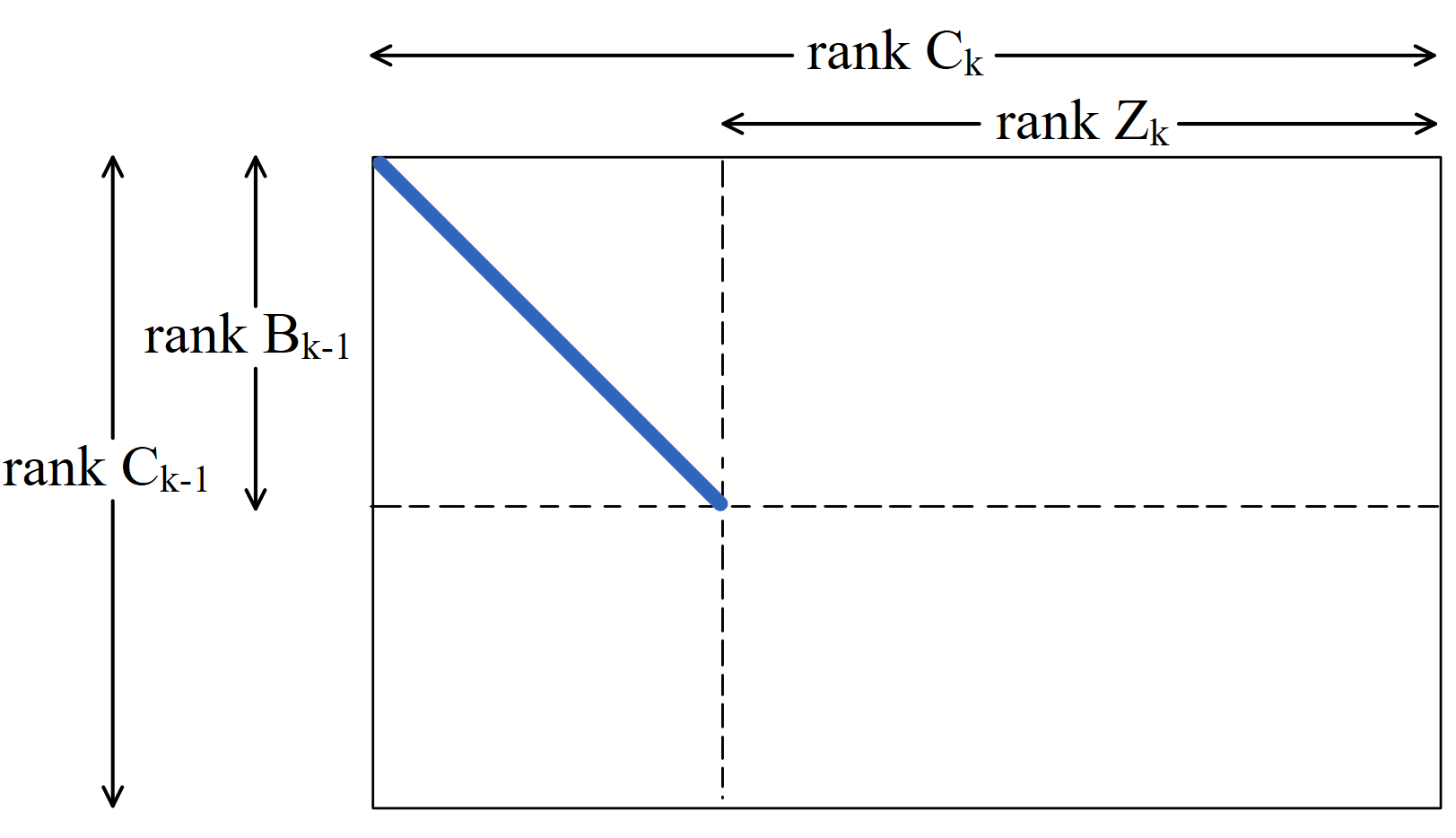}
  \caption{Interpreting the Smith Normal Form of the boundary matrix
    $\partial_k$. The blue line represents a diagonal of ones, all other
    elements being zero. Based on Figure IV.5 in \cite{edelsbrunner_computational_2010}.}
  \label{smith}
\end{figure}

\noindent
The boundary matrix, reduced to its SNF, takes the block diagonal form in Figure \ref{smith}. In particular, the number of zero columns in the SNF of the boundary matrix counts the $\text{rank}(\mathbb{Z}_k)$ and the number of nonzero rows gives $B_{k-1}$.

\noindent
To see why, recall that $B_{k-1}$ is the image of the boundary of $C_k$, i.e. $B_{k-1} =
\partial_k c$, with $\ c \in C_k$. The rank of $B_{k-1}$ is then the number of linearly
independent rows of $\partial_k$, as is shown in Figure \ref{smith}. On the
other hand, since $Z_k$ is the kernel of $\partial_{k}$ acting on $C_k$, the rank of
$Z_k$ is the difference between the rank of $C_k$ and the rank of the linearly independent columns of
$\partial_k$. Now we have all we need to compute Betti numbers. For example, let's calculate $b_1$ of $X$. 
The Smith normal form of $\partial_{1}(X)$ is given by
\begin{equation*}
  \mathrm{SNF}(\partial_1) =
    \begin{pmatrix}
      1 & 0 & 0 & 0 & 0 & 0 \\
      0 & 1 & 0 & 0 & 0 & 0 \\
      0 & 0 & 1 & 0 & 0 & 0 \\
      0 & 0 & 0 & 1 & 0 & 0 \\
      0 & 0 & 0 & 0 & 0 & 0 \\
    \end{pmatrix}.
\end{equation*}
The two zero columns tell us that $\text{rank}(Z_1) = 2$. To obtain $B_1$, we
need to look at the boundary matrix $\partial_2$:
\begin{equation*}
  \partial_2 = 
  \begin{blockarray}{cc}
    & \langle a,b,c \rangle \\
     \begin{block}{c(c)}
      {\langle a,b \rangle}& 1 \\
      {\langle b,c \rangle}& 0 \\
      {\langle c,d \rangle}& 0 \\
      {\langle d,e \rangle}& 0 \\
      {\langle e,a \rangle}& 1 \\
      {\langle b,e \rangle}& 1 \\
     \end{block}
 \end{blockarray}.
\end{equation*}
The Smith Normal Form of $\partial_2$ is obviously the vector with a 1 in its
first position and 0s everywhere else. Since there is one non-zero column, $\text{rank}(B_1) = 1$ and finally
\begin{equation*}
   b_1 = \text{rank}(Z_1) -\text{rank}(B_1) = 2 - 1 = 1,
\end{equation*}
The other Betti numbers are easily obtained in the same manner. The advantage of this technique is that it relies solely on the Gauss-reduction of a sparse matrix which is fast to compute, even for the large matrices encountered in data analysis problems. 

\subsection{Persistent homology}

Recall from our discussion in the introduction that there is one important parameter in the 
construction of a simplicial complex from data: the scale parameter $\epsilon$. If $\epsilon$ is small, the
complex will just be the set of all the data points; if $\epsilon$ is large on the other hand, the
complex will be a high dimensional complex built on a connected nearest
neighbour graph. Obviously, we want the conclusions we draw to depend only on the data, not our choice of
parameters. Hence, we need to perform the analysis on a range of values of $\epsilon$, to extract those topological 
features that \textit{persist} over a range of scales.\\

\noindent
The na\"{\i}ve approach to this problem of \textit{persistent homology} is to choose a
discrete set of $\epsilon$ values, then calculate all the Betti numbers for each
$\epsilon$. This would be mind-numbingly slow to compute for even moderately sized datasets,
essentially because of the numerous simplices that the complex will contain when $\epsilon$
is large. \\

\noindent
Fortunately, there is an elegant algorithm for computing Betti numbers over a
range of scale values, which requires only one matrix reduction to be performed.
The trick is that if we make $\epsilon$ large, we obtain a maximal simplicial
complex, from which the homology of all complexes at scales smaller than $\epsilon$ can be 
computed. \\

\begin{figure}[ht!]
  \centering \includegraphics[height=4cm,width=\textwidth]{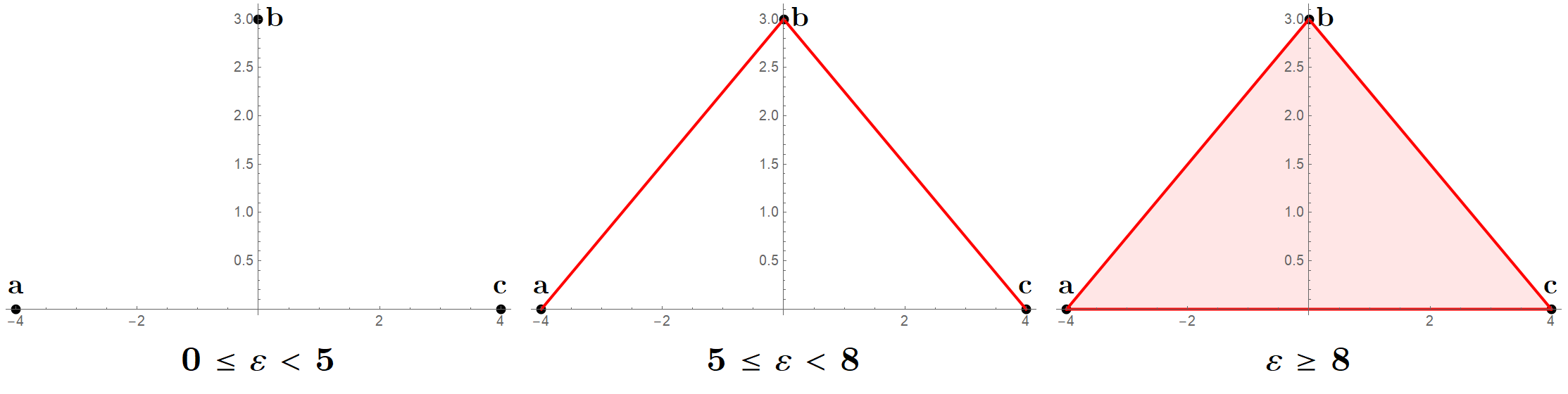}
  \caption{Geometric representation of $Y(\epsilon)$ as $\epsilon$ is varied.}
  \label{triangle_f}
\end{figure}

\noindent
A trivial example will suffice to illustrate this discussion. Consider a dataset
of three observations: $D = \{a = (4,0), b= (-4,0),$ and $c =(0,3)\}$. Using Pythagoras' theorem, 
we see that the nearest neighbour graph of D will be a totally disconnected for $\epsilon < 5$, 
a cycle for $5 \leq \epsilon < 8$, and a connected graph for
$\epsilon \geq 8$. As we can see in Figure \ref{triangle_f}, the corresponding
Vietoris-Rips complexes ($Y(\epsilon)$) are $Y(0) = \{a,b,c\}$, $Y(5) =
\{a,b,c,\langle a,b \rangle,\langle b,c \rangle\}$ and $Y(8) = \{a,b,c,\langle a,b \rangle,\langle b,c \rangle,\langle c,a \rangle,\langle a,b,c \rangle\}$. \\

\noindent
It helps to think of Figure \ref{triangle_f} as a construction process: 
\begin{itemize}
   \item Start with $Y(0)$, and add simplices whenever $\epsilon$ is large enough for them to appear, due to a change in the nearest-neighbour graph. 
   \item Continue in this manner, until we add the highest dimensional simplex possible (in this case, $\langle a,b,c \rangle$). 
   \item Choose a maximal complex, in this case $Y = Y(8)$.
\end{itemize}

\noindent
It is useful to consider this construction as a \textit{filtration} of some monotonic
function $f: Y \rightarrow \mathbb{R}$. If we sort the simplices of
$Y$ using $f$, the resulting sequence should describe the order by which
simplices are added to $Y(0)$, as $\epsilon$ increases to the value corresponding
to the maximal simplicial complex (in this case $\epsilon = 8$). 

\begin{table}[ht!]
  \centering
  \begin{tabular}{|c|c|c|c|c|c|c|c|}
    \hline
    Index $(i)$      & 1 & 2 & 3& 4 & 5 & 6 & 7 \\
    \hline
    Simplex ($\sigma_i$) & $\langle a \rangle $ & $\langle b \rangle $ & $\langle c \rangle $ & $\langle a,b \rangle $ & $\langle b,c \rangle $ & $\langle c,a \rangle $ & $\langle a,b,c \rangle $ \\
    \hline
    $f(\sigma_i)$        & 0   & 0.1 & 0.2 &   5   &  5.1  &   8   &  8.1 \\
    \hline
    $\epsilon_i$     & 0 & 0 & 0 & 5 & 5 & 8 & 8  \\                                      
    \hline
  \end{tabular}
  \caption{A possible filtration function for $Y$ which can be used to
    produce Figure \ref{triangle_f}. $\epsilon_i$ indicates the value of
    $\epsilon$ for which simplex $\sigma_i$ first appears in the complex.}
 \label{triangle_t}
\end{table}

\subsubsection{Filtrations}

Since they will be important in what follows, let's talk a little more about filtrations. 
In general, if $K$ is simplicial complex with $n$ simplices, a
\textit{filtration} is a cover of $K$ given by $K_{\epsilon}$ where $\epsilon \in T \subseteq \mathbb{R}$, and 
$\epsilon \leq \delta \Rightarrow K_{\epsilon} \subseteq K_{\delta}$ \cite{chazal_introduction_2017}.\\

\noindent
A \textit{sublevel set filtration} of $K$ is a special case of a filtration,
where we use a monotonic function $f: K \rightarrow \mathbb{R}$ to form the
cover of $K$ given by $(f^{-1}(-\infty, \epsilon])_{\epsilon \in \mathbb{R}}$
\cite{edelsbrunner_computational_2010}. Notice that there are finitely many
distinct complexes $K(\epsilon) = f^{-1}(-\infty, \epsilon])$. In fact, there are at most
$m + 1 \leq n+1$ of them,
\begin{equation*}
K(\epsilon_0) = \emptyset \subseteq K(\epsilon_1) \subseteq K(\epsilon_2) \subseteq \dots \subseteq K(\epsilon_m) =
K. 
\end{equation*}
We want a general method for finding a function $f$ as in Table \ref{triangle_t}, by which we can impose total order on a simplicial complex $K$, corresponding to the order by which simplices are added as $\epsilon$ increases. For all $\sigma \subseteq K$ with $\sigma \neq \emptyset$, let $V(\sigma)$ be the set of vertices of $\sigma$ and $E(\sigma)$ be the set of edges which are faces of $\sigma$, if they exist. If, in addition, we denote the length of an edge $\alpha$ by $l(\alpha)$ and label the vertices as $V(K) = \{0, 1, \dots, k\}$, then, the filtration function $f$ is chosen to satisfy that, for any two simplices $\sigma,\tau \in K$,

\begin{enumerate}
  \item $f$ is decreases along faces, i.e. $\text{dim}(\sigma) <
    \text{dim}(\tau) \Rightarrow f(\sigma) < f(\tau)$.
  \item If the two simplices have the same dimension then we compare the lengths of edges. In particular if, 
  \begin{equation*}
    \text{dim}(\sigma) = \text{dim}(\tau) \text{ and }
    \max_{\gamma \in E(\sigma)}l(\gamma) < \max_{\gamma \in E(\tau)}l(\gamma) \Rightarrow f(\sigma) < f(\tau).
  \end{equation*}
  \item If $\text{dim}(\sigma) = \text{dim}(\tau)$ and $\max_{\gamma \in E(\sigma)}l(\gamma) = \max_{\gamma \in E(\tau)}l(\gamma)$, then we compare the maximum of
      the vertex numbers of $\sigma$ and $\tau$ which are not shared between
      them. Specifically, if $W = V(\sigma)\cap V(\tau)$, then
\begin{equation*}
\max\{v \in V(\sigma) \backslash W\} <  \max\{v \in V(\tau) \backslash W\}   
      \Rightarrow f(\sigma) < f(\tau).
    \end{equation*}
\end{enumerate}
This allows us to construct a monotonically decreasing filtration function of the scale parameter $\epsilon$ in much the same spirit as Zamolodchikov's c-function in 2-dimensional conformal field theories. This is no coincidence. The c-function and its associated c-theorem are one of the most important results in modern renormalization group theory \cite{Zamolodchikov:1986gt} and, in some sense, filtrations track the course-graining of the data set from small to large scales. It would be of enormous interest to explore the connection between renormalization group theory and this course-graining of data further.

\subsubsection{Homology of filtrations}
Using the filtration $f$ from the previous section, let's see now how
the homology of $K(\epsilon) = f^{-1}(-\infty, \epsilon]$ changes as $\epsilon$
is varied. This will serve to unpack some of the details of our triangle example in Fig. \ref{triangle_f}.
We start by noticing that $Y$ goes from having three simplices in $H_0$ to just one at $\epsilon =
5$. In other words, two of the connected components `die' at $\epsilon = 5$, {\it i.e.} they are merged into one piece.\\

\begin{figure}[ht!]
  \centering
  \subfloat[Barcode diagram]{{\includegraphics[width=6.5cm]{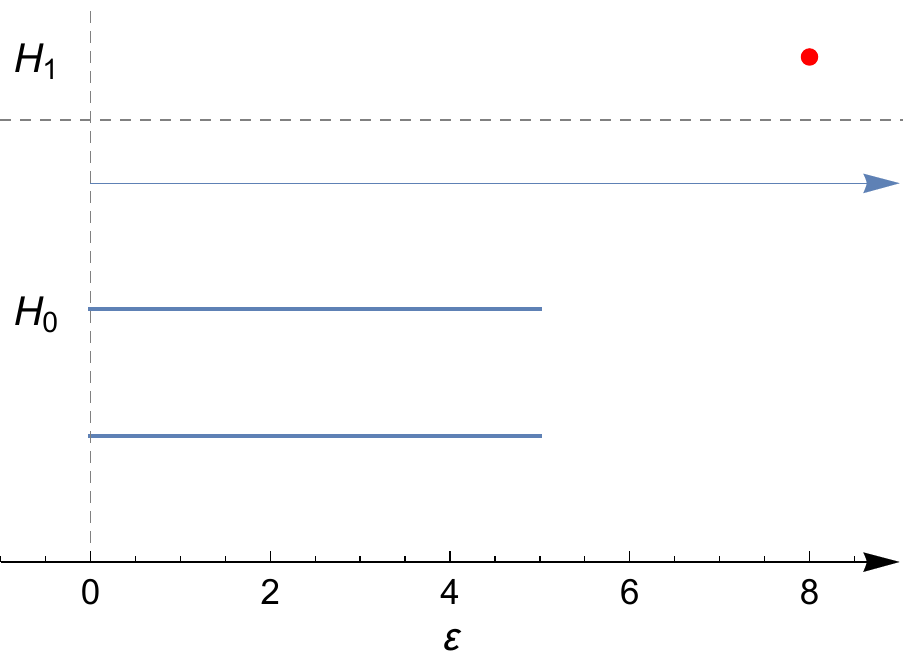}}}
  \quad
  \subfloat[Persistence diagram]{{\includegraphics[width=6.5cm]{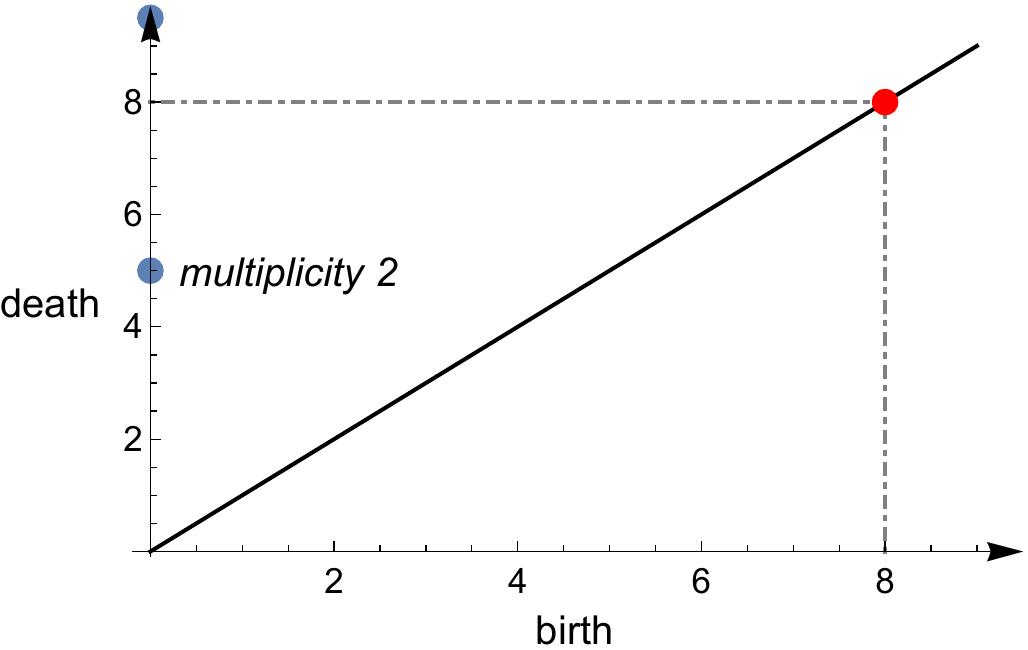}}}
  \caption{Persistent homology diagrams of the simplicial complex Y.}
  \label{triangle_pers}
\end{figure}

\noindent
When $\epsilon = 8$, a new feature in $H_1$ is `born'. This is the cycle 
$\langle a,b \rangle + \langle b,c \rangle + \langle c,d \rangle$, which immediately dies when we colour it in to
form the 2-simplex $\langle a,b,c \rangle$. We can represent these changes in homology with
two equivalent plots, known as the \textit{persistence diagram} and
\textit{barcode diagram}, displayed in Figure \ref{triangle_pers}. In Figure \ref{triangle_pers}b, 
the colours indicate which homology class the points in the diagram are related to. In general, we denote by
$\text{Dgm}_p(f)$ the multiset containing the points in the persistence
diagram of $H_p$. In our example, $\text{Dgm}_0(f) = \{\{0,5\},\{0,5\},\{0,\infty\}\}$
and $\text{Dgm}_1(f) = \{\{8,8\}\}$. \\

\noindent
Persistence and barcode diagrams provide a simple visual way to find the Betti numbers of
$K(\epsilon)$ for any $\epsilon$. For example, the three bands in the $H_0$
band of the barcode diagram in Figure \ref{triangle_pers}, tell us that $\text{dim} H_0 = b_0 = 3$ at
this $\epsilon$ value. Notice that the longer a band of the barcode is, the more
persistent the corresponding feature is in this filtration. Equivalently, points in the
persistence diagram which are furthest away from the birth = death line
correspond to the most persistent features.

\subsubsection{Computing persistence}

For any data set $M$, we can construct a general, computationally efficient method to find barcode or persistence
diagrams according to the following algorithm:

\begin{enumerate}
\item Choose a `large' value of $\epsilon$ and construct the VR complex  $\mathcal{V}_\epsilon(M)$, using method described in, for example \cite{zomorodian_fast_2010}.
\item Next, sort the simplices of $\mathcal{V}_\epsilon(M)$ according to their corresponding
  values given by the filtration function $f$, as described in the previous
  section.
\item Construct the boundary matrix $\partial(\mathcal{V})$.
\item Let low($j$) be the row index of the lowest 1 of column $j$ in
  $\partial(\mathcal{V})$, with low($j) \equiv 0$ if $j$ is a column of zeroes. With the index defined, column-reduce 
  $\partial(\mathcal{V})$ without exchanging columns: {\it i.e.} moving from $j = 1$ to $n$, add column $k < j$ to
  column $j$ whenever $\text{low}(j) = \text{low}(k)$. Let $C$ be the resulting
  column-reduced matrix.
\item If $j$ is a column of zeros in $C$, then the addition of the simplex
  $\sigma_j$ results in the birth of a new cycle. If, on the other hand, there is an $i > j$ such that
  $\text{low}(i) = j$, then the cycle dies when simplex $\sigma_i$ is added. If
  no such $i$ exists, the cycle never dies.
\end{enumerate}
\noindent
Let's elaborate on the last two steps of the algorithm with our usual example.
After an initial sort of the simplices in $\mathcal{V}_\epsilon(X)$ as per Table
\ref{triangle_t}, the boundary matrix of $Y$ is given by

\begin{equation*}
  \partial Y = 
  \begin{blockarray}{ccccccc}
    \langle a \rangle & \langle b \rangle & \langle c \rangle & \langle a,b \rangle & \langle b,c \rangle & \langle c,a \rangle & \langle a,b,c \rangle \\
    \begin{block}{(ccccccc)}
      0 & 0 & 0 & 1 & 0 & 1 & 0 \\
      0 & 0 & 0 & 1 & 1 & 0 & 0 \\
      0 & 0 & 0 & 0 & 1 & 1 & 0 \\
      0 & 0 & 0 & 0 & 0 & 0 & 1 \\
      0 & 0 & 0 & 0 & 0 & 0 & 1 \\
      0 & 0 & 0 & 0 & 0 & 0 & 1 \\
      0 & 0 & 0 & 0 & 0 & 0 & 0 \\
    \end{block}
  \end{blockarray}\,\,.
\end{equation*}

\noindent
Proceeding from left to right, we find low$(j)$ for each column $j$, and check
if any previous column $k$ satisfies low$(k) = \text{low}(j)\neq0$. There is no
column $j < 6$ which satisfies this condition. However, $\text{low}(6) = \text{low}(5)$, so we
add column 5 to column 6:
\begin{equation*}
  \begin{blockarray}{ccccccc}
    \langle a \rangle & \langle b \rangle & \langle c \rangle & \langle a,b \rangle & \langle b,c \rangle & \pbox{20cm}{\ \ \ $\langle c,a \rangle$ \\ +$\langle b,c \rangle$} & \langle a,b,c \rangle \\
    \begin{block}{(ccccccc)}
      0 & 0 & 0 & 1 & 0 & 1 & 0 \\
      0 & 0 & 0 & 1 & 1 & 1 & 0 \\
      0 & 0 & 0 & 0 & 1 & 0 & 0 \\
      0 & 0 & 0 & 0 & 0 & 0 & 1 \\
      0 & 0 & 0 & 0 & 0 & 0 & 1 \\
      0 & 0 & 0 & 0 & 0 & 0 & 1 \\
      0 & 0 & 0 & 0 & 0 & 0 & 0 \\
    \end{block}
  \end{blockarray}\,\,,
\end{equation*}
where we recall that we are still counting modulo 2. Now column 4 and
column 6 are identical, so we add column 4 to column 6 to obtain

\begin{equation*}
  C(Y) = 
  \begin{blockarray}{ccccccc}
    \langle a \rangle & \langle b \rangle & \langle c \rangle & \langle a,b \rangle & \langle b,c \rangle & \pbox{20cm}{\quad $\langle c,a \rangle$ \\ +$\langle b,c \rangle$ \\ +$\langle a,b \rangle$} & \langle a,b,c \rangle \\
    \begin{block}{(ccccccc)}
      0 & 0 & 0 & 1 & 0 & 0 & 0 \\
      0 & 0 & 0 & 1 & 1 & 0 & 0 \\
      0 & 0 & 0 & 0 & 1 & 0 & 0 \\
      0 & 0 & 0 & 0 & 0 & 0 & 1 \\
      0 & 0 & 0 & 0 & 0 & 0 & 1 \\
      0 & 0 & 0 & 0 & 0 & 0 & 1 \\
      0 & 0 & 0 & 0 & 0 & 0 & 0 \\
    \end{block}
  \end{blockarray}\,\,.
\end{equation*}
Now column 6 is all zeroes, corresponding to a cycle, whose boundary is
always zero. This tells us that the addition of the simplex $\sigma_6 = \langle c,a \rangle$ causes the birth
of a cycle. This is true in general; for any simplicial complex $K$, if the
$j^{\text{th}}$ column of its column-reduced matrix is zero, then the addition
of simplex $\sigma_j$ induces the birth of a cycle.\\

\noindent
Next, notice that  low$(7) = 6$ in $C(Y)$. This means that the simplex $\sigma_7 =
\langle a,b,c \rangle$
has a boundary corresponding to the cycle accumulated in column 6, such that the
cycle dies when $\langle a,b,c \rangle$ is added to the complex. Again this is a consequence
of the general property of such a reduced matrix; for any non-zero column $j$
with low($j) = i$, the addition of the simplex $\sigma_j$ causes the death of
the cycle which was born when $\sigma_i$ was added. \\

\noindent
To read off the persistence intervals from $C(Y)$, we first find
$[\text{low}(j),j]$ for all the non-zero columns $j$. The elements of this pair
are the indices of the $\epsilon$ values corresponding to the birth and death of a
cycle, respectively. The dimension of this cycle is that of the simplex which
triggered its formation, namely dim($\sigma_{low(j)}$). In our case, we have the 0-dimensional index intervals $[2,4],
[3,5]$ and a 1-dimensional index interval $[6,7]$. Looking up the corresponding $\epsilon$ value for each index in Table
\ref{triangle_t}, we obtain the intervals $[\epsilon_2, \epsilon_4] = [0,5],
[\epsilon_3,\epsilon_5] = [0,5]$ for 0-cycles and  $[\epsilon_6,\epsilon_7] =
[8,8] = \{8\}$ for the 1-cycle.\\

\noindent
Infinite persistence intervals are of the form $[\epsilon_i,\infty]$, where
column $i$ is zero (a cycle is born there), but with no $j$ such that $i =
\text{low}(j)$ (the cycle never dies). There is only one in this case; the index
interval $[1,\infty]$, which corresponds to $[\epsilon_1,\infty] = [0,\infty]$.
Putting this all together, we have $\text{Dgm}_0(f) = \{\{0,5\},\{0,5\},\{0,\infty\}\}$ and
$\text{Dgm}_1(f) = \{8,8\}$, corroborating Figure \ref{triangle_pers}.

\subsubsection{A less trivial example}

\begin{figure}[ht!]
  \centering \includegraphics[height=4cm,width=5cm]{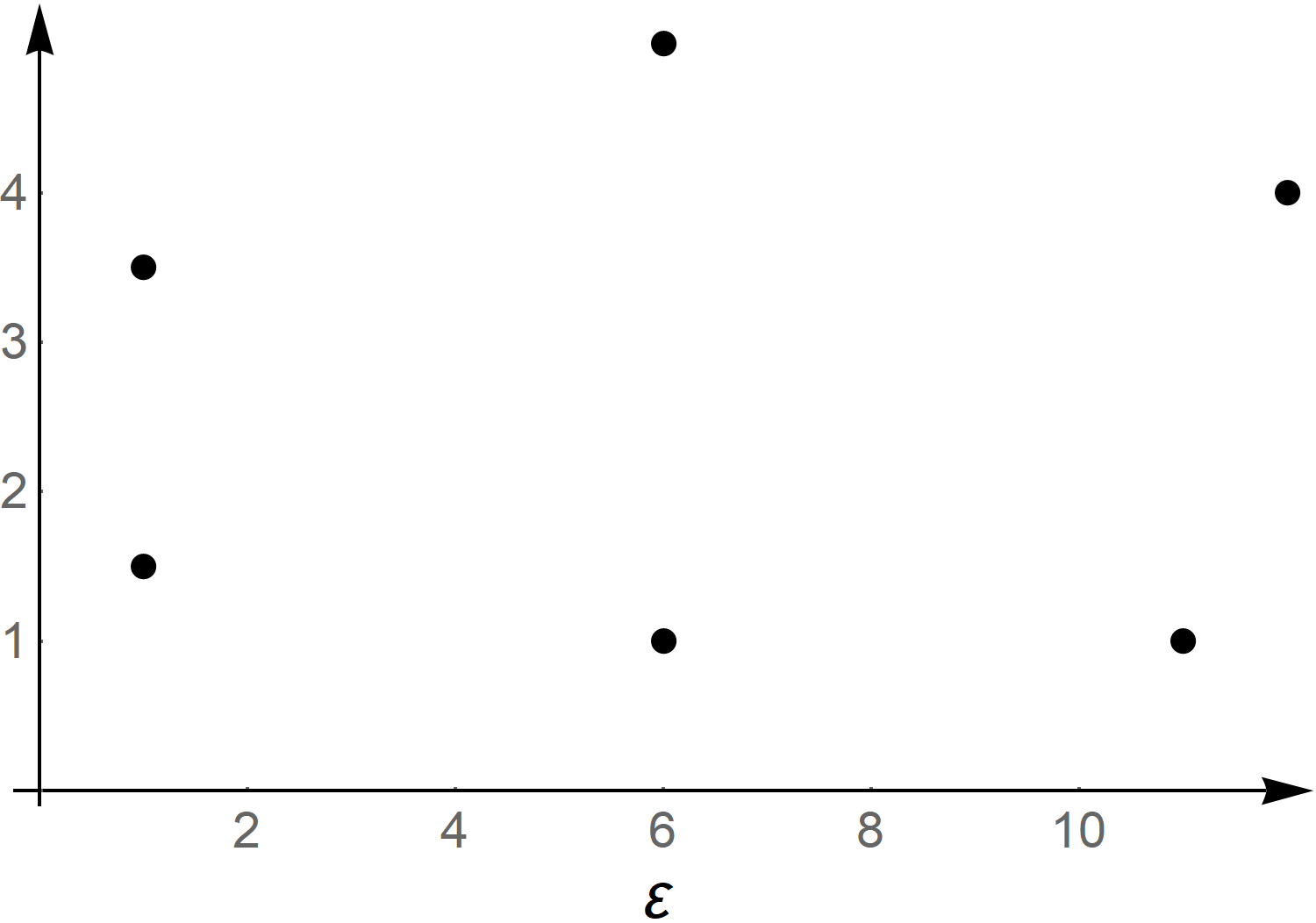}
  \caption{A small dataset $D$ in $\mathbb{R}^2$.}
  \label{boxpoints}
\end{figure}

\begin{figure}[ht!]
  \centering \includegraphics[width=7.5cm]{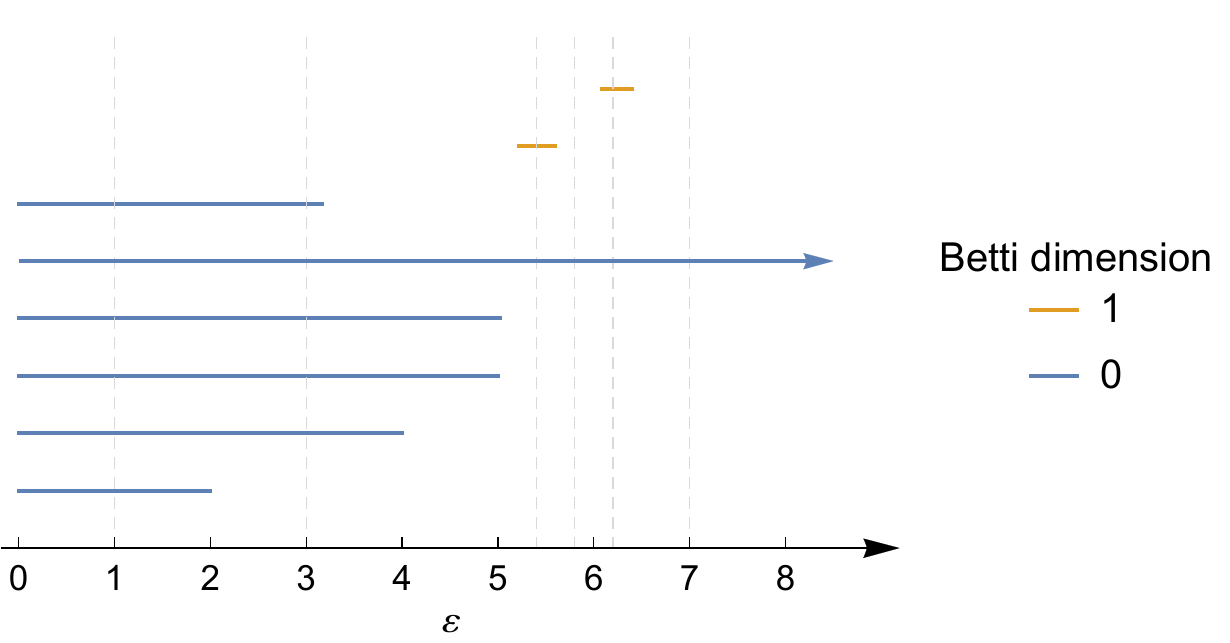}
  \caption{Barcode diagram for the dataset $D$. The dashed grey lines indicate
    the values of $\epsilon$ used in constructing Figure \ref{boxbetti}.}
  \label{boxbarcode}
\end{figure}

Of course, our standard exemplar complex $X$ is, by design, fairly trivial. To illustrate this construction in a slightly less trivial setting, consider now the small dataset $D \subset \mathbb{R}^2$,  pictured in Figure \ref{boxpoints}.
The barcode for $D$ as $\epsilon$ ranges from 0 to 8 is depicted in Figure \ref{boxbarcode}. We see that the first edge in the VR complex of the data appears at $\epsilon = 2$, after which more edges form, gradually reducing the number of separate components of the complex until only one remains at $\epsilon = 5$. A 1-dimensional hole appears briefly around $\epsilon = 5.4$ (represented by the orange line), and dies soon afterwards as it replaced by two 2-simplices. Another hole appears around $\epsilon = 6.2$ and  no further holes manifest for larger values of $\epsilon$.\\

\begin{figure}[ht!]
  \centering \includegraphics[height=10cm,width=8cm]{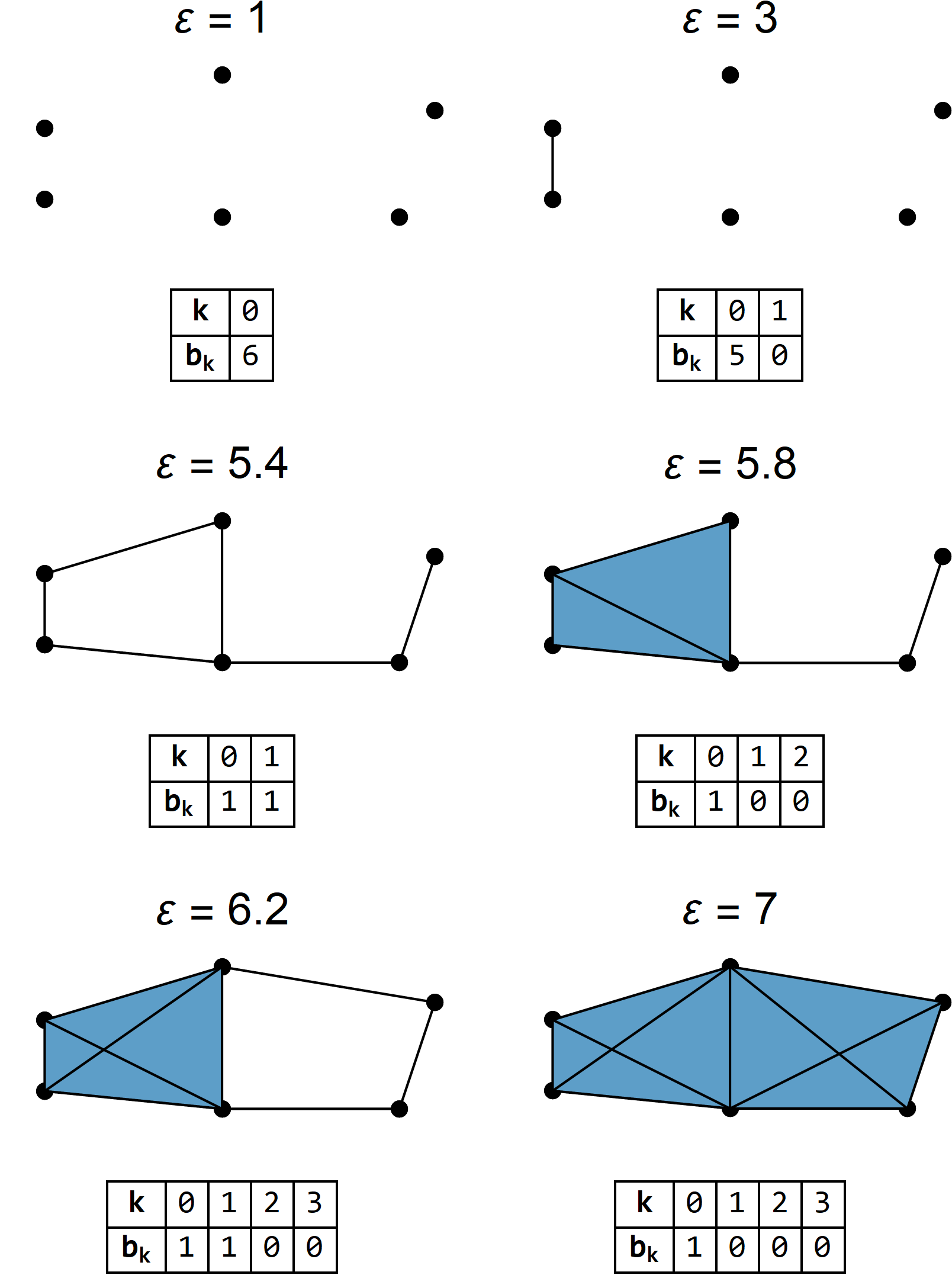}
  \caption{Evolution of the VR-complex of the dataset $D$, as $\epsilon$
    increases. $b_k$ is the $k^{th}$ Betti number of the corresponding complex.}
  \label{boxbetti}
\end{figure}
\noindent
In this simple case, we can check our conclusions from the barcode diagram by directly computing Betti numbers for relevant values of $\epsilon$, using the technique described in Section \ref{bettimatrix}. The result is shown in Figure \ref{boxbetti}. Indeed, we now see the two holes which we found in the barcode diagram, for relevant values of $\epsilon$. Furthermore, we conclude that no more interesting topological features occur once $\epsilon$ reaches 7. 

\clearpage
\section{The Mapper Algorithm}

\subsection{Fundamentals}
Whereas persistent homology is concerned with the number of `holes'
in the shape of data at different scales, the Mapper algorithm is concerned with visualising the shape of data
through a particular lens. More than just a figure of speech, the lens or filter is a function which maps the data
to $\mathbb{R}^n$, usually just $\mathbb{R}$, chosen to reveal useful
information about the dataset.

\begin{figure}[ht!]
  \centering \includegraphics[width=\textwidth]{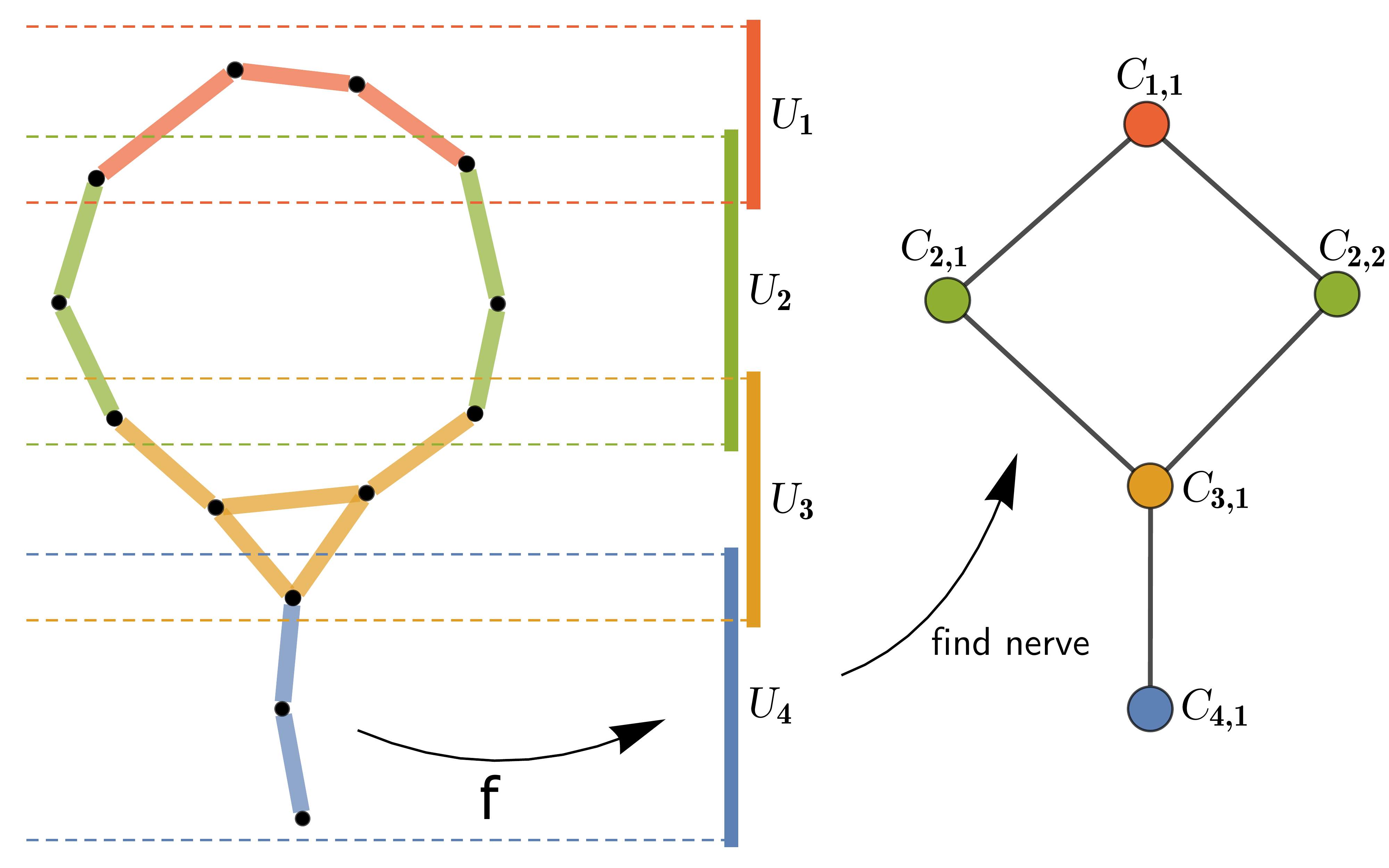}
  \caption{Illustrative example of the Mapper algorithm, as applied to a simple
    2D dataset (the black dots) using the height function as the filter ($f$). The nearest
    neighbourhood graph was used to cluster the pullback cover, as shown in the
    leftmost plot.}
  \label{Mapperex1}
\end{figure}
\noindent Mapper is perhaps best illustrated visually, as with the example in Figure
\ref{Mapperex1}. The basic steps of the method are as follows \cite{chazal_introduction_2017}:
\begin{enumerate}
\item Map the dataset $M$ to $\mathbb{R}$ using a filter function $f$.
  Common choices for such a filter function include the real-valued {\it centrality} and {\it eccentricity} functions
  $f_{\mathrm{c}}(x) = \sum_{y\in M}d(x,y)$ and $f_{\mathrm{e}}(x) = \max_{y\in M}d(x,y)$, respectively.

\item Construct a \textit{cover} of the filter values, $\mathcal{U} = (U_i)_{i \in
    I}$, usually in the form of a set of overlapping intervals, as pictured in
  the middle of Figure \ref{Mapperex1}. Unlike in this example, the intervals are usually chosen
  to be of constant length $r$ and constant overlapping percentage $g$. These are known as
  the \textit{resolution} and \textit{gain} of the cover, respectively.

\item Find the \textit{pull back cover} of $M$ induced by
$(f,\mathcal{U})$. This is the collection of sets $\big(f^{-1}(U_i)\big)_{i \in
  I}$. In Figure \ref{Mapperex1}, there are four sets ($U_1,U_2,U_3$ and $U_4$) in the
cover, each having a preimage given by all the points between the dotted lines
of that interval. 

\item For each $U_i \in \mathcal{U}$, cluster $f^{-1}(U_i)$ into sets $C_{i,1}, ...,
  C_{i,k_i}$. The resulting collection of clusters is called the \textit{refined
    pullback cover}: $(C_{i,j})_{i \in I, j \in \{1, .., k_i\}}$. The choice of
  clustering method is up to the analyst. In this paper, we follow the
    suggestion of \cite{chazal_introduction_2017} and cluster $f^{-1}(U_i)$ in to
  the connected components of the subgraph of an $\epsilon$-neighbouring graph of $M$,
  with vertices $f^{-1}(U_i)$.

In our example, the only element of the pullback cover with more than one
cluster corresponds to the green interval. We see that the green interval's preimage includes
points on two opposite sides of the circle, such that the subgraph which they
generate (with edges coloured green) is disconnected. Hence, the refined
pullback cover has five elements: $C_{1,1},C_{2,1},C_{2,2},C_{3,1},C_{4,1}$.  

\item Construct the \textit{nerve of the refined pull back cover}, which is
  a simplicial complex in general, and which reduces to a graph if the cover is
  chosen such that no more than two intervals overlap at any point. This graph
  has vertices $(C_{i,j})_{i \in I, j \in \{1, .., k_i\}}$, and an edge between
  any $C_{i,j}$ and $C_{k,l}$ if and only if $C_{i,j} \cap C_{k,l} \neq
  \emptyset$. This is the final product of the Mapper algorithm, as given be the
  rightmost graph in our example. 

\end{enumerate}
In this way, given a particular data set, Mapper efficiently allocates to it a simplicial complex. The steps outlined above are algorithmic and easily implemented in the reader's favourite language. 
\subsection{Applications of Mapper to 3D point data}

For the purpose of illustrating how it works on, debatably, more realistic data sets, we implemented the Mapper algorithm in \texttt{Mathematica}, using the clustering method of connected components of the $\epsilon$-neighbouring graph. The algorithm was then applied to point data sampled from various 3D objects, as obtained from CAD files, with the goal of uncovering the topology of the object from the data points alone. All 3D objects were standardised to have height 1 in $\mathbb{R}^3$.

\begin{figure}[ht!]
  \centering
  \subfloat{{\includegraphics[width=5cm]{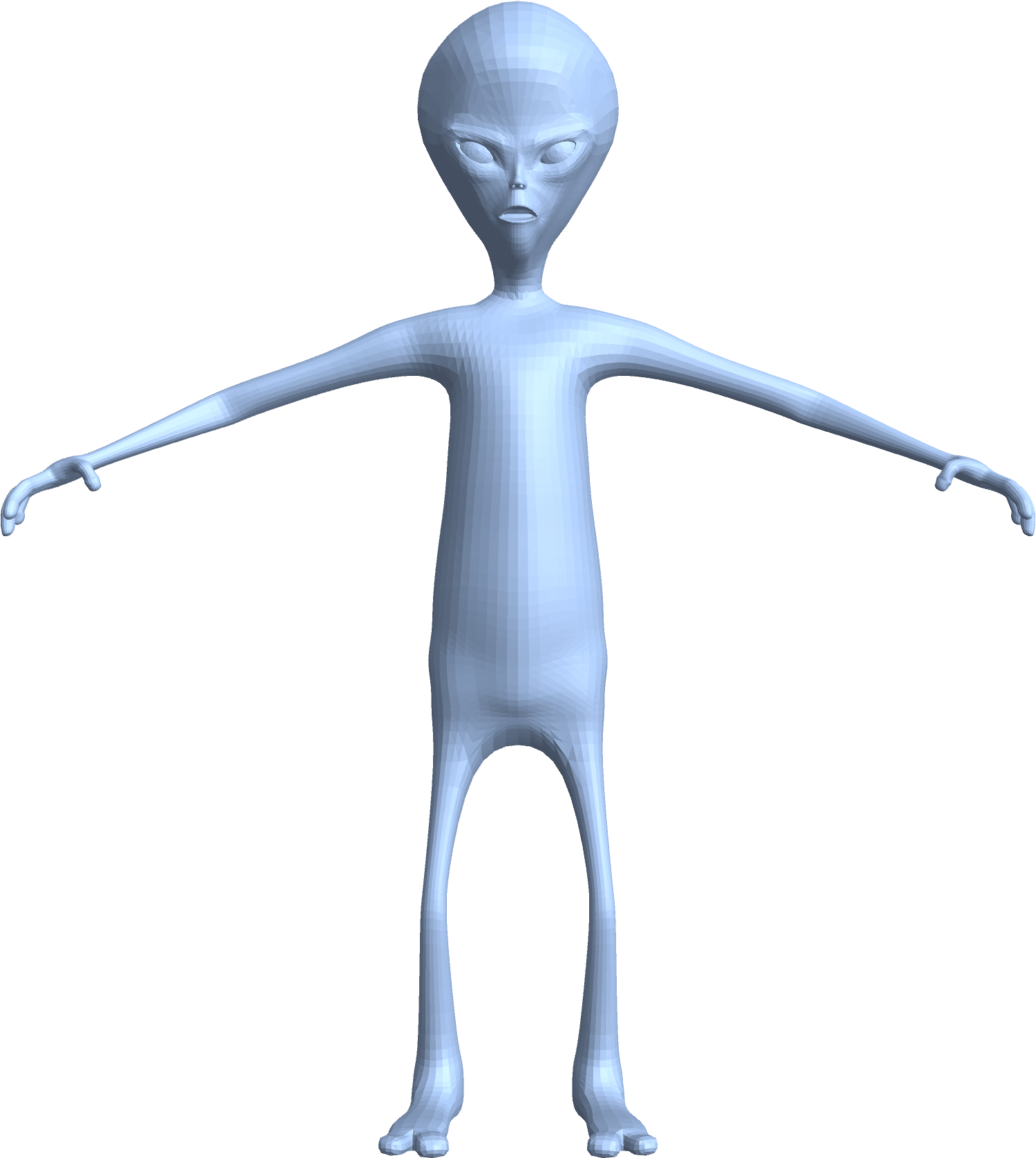}}}
  \quad
  \subfloat{{\includegraphics[width=5cm]{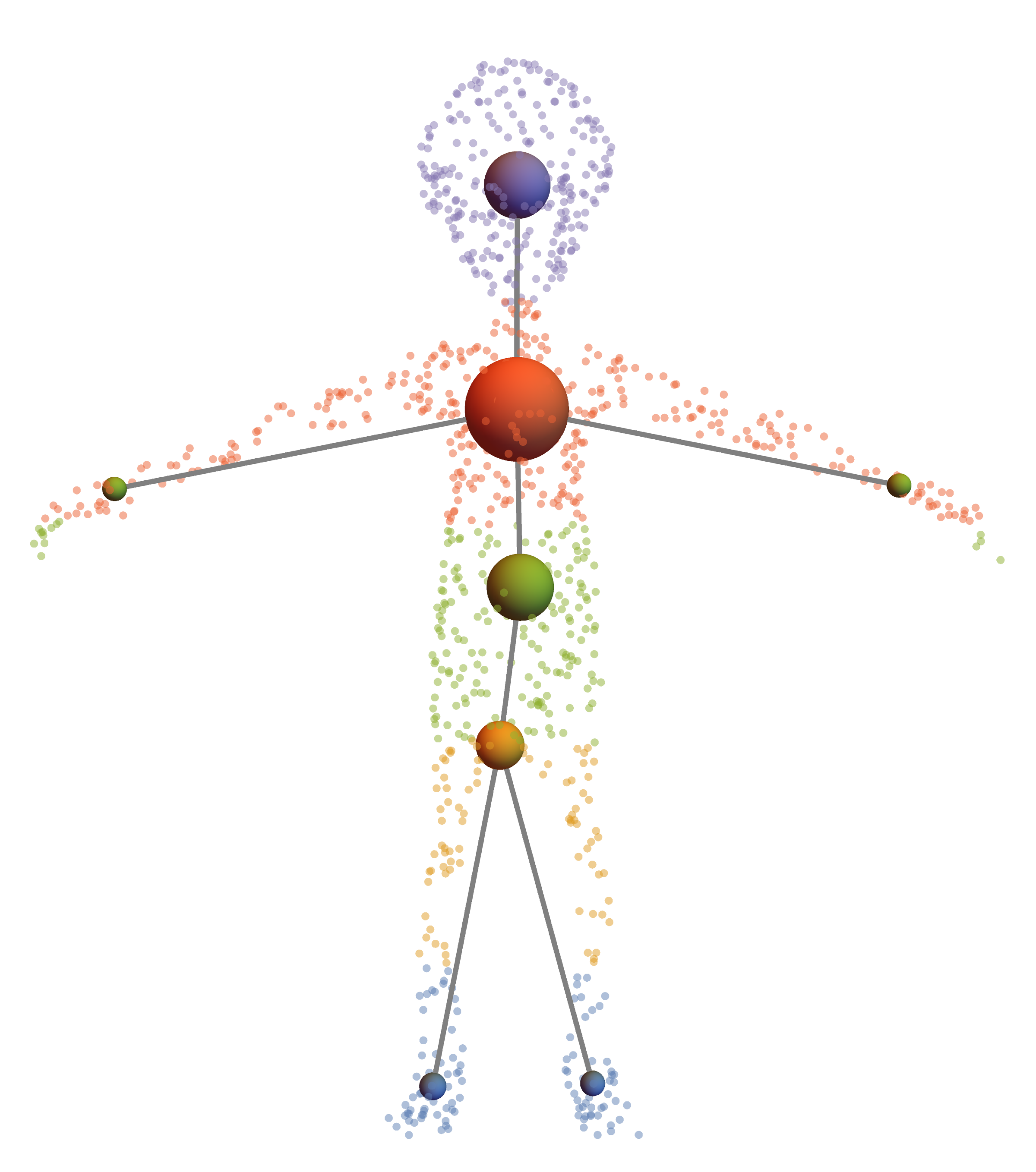}}} \\
  \subfloat{{\includegraphics[width=9cm]{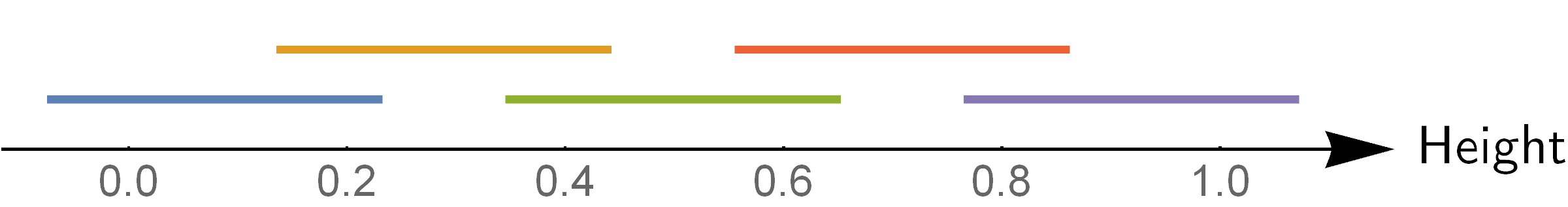}}}
  \caption{1000 points were sampled uniformly from the 3D model of an alien\cite{printable_models_grayalien_nodate} (left), then
    filtered by height to obtain the graph (right), which is overlayed over the
    data. The points and graph vertices are coloured by the interval in the
    cover (bottom) to which they correspond to.}
  \label{alien1}
\end{figure}

In the first example, points sampled from an alien figure were filtered by
height, to obtain Figure \ref{alien1}. Notice that graph output of Mapper
identifies the core structure of the object: a mass with five appendages (head,
legs and arms). 

\begin{figure}[ht!]
  \centering
  \subfloat{{\includegraphics[width=5cm]{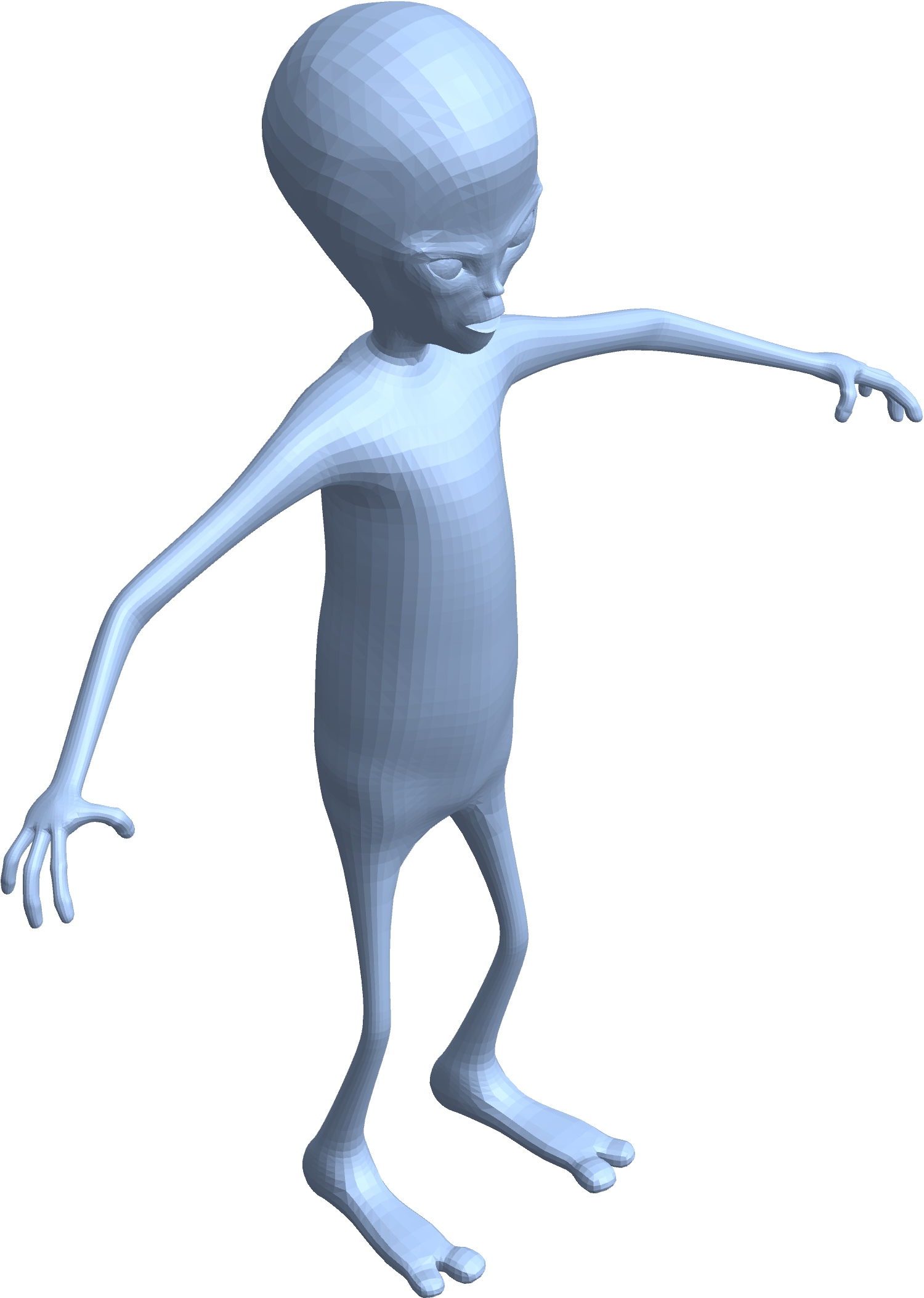}}}
  \quad
  \subfloat{{\includegraphics[width=5cm]{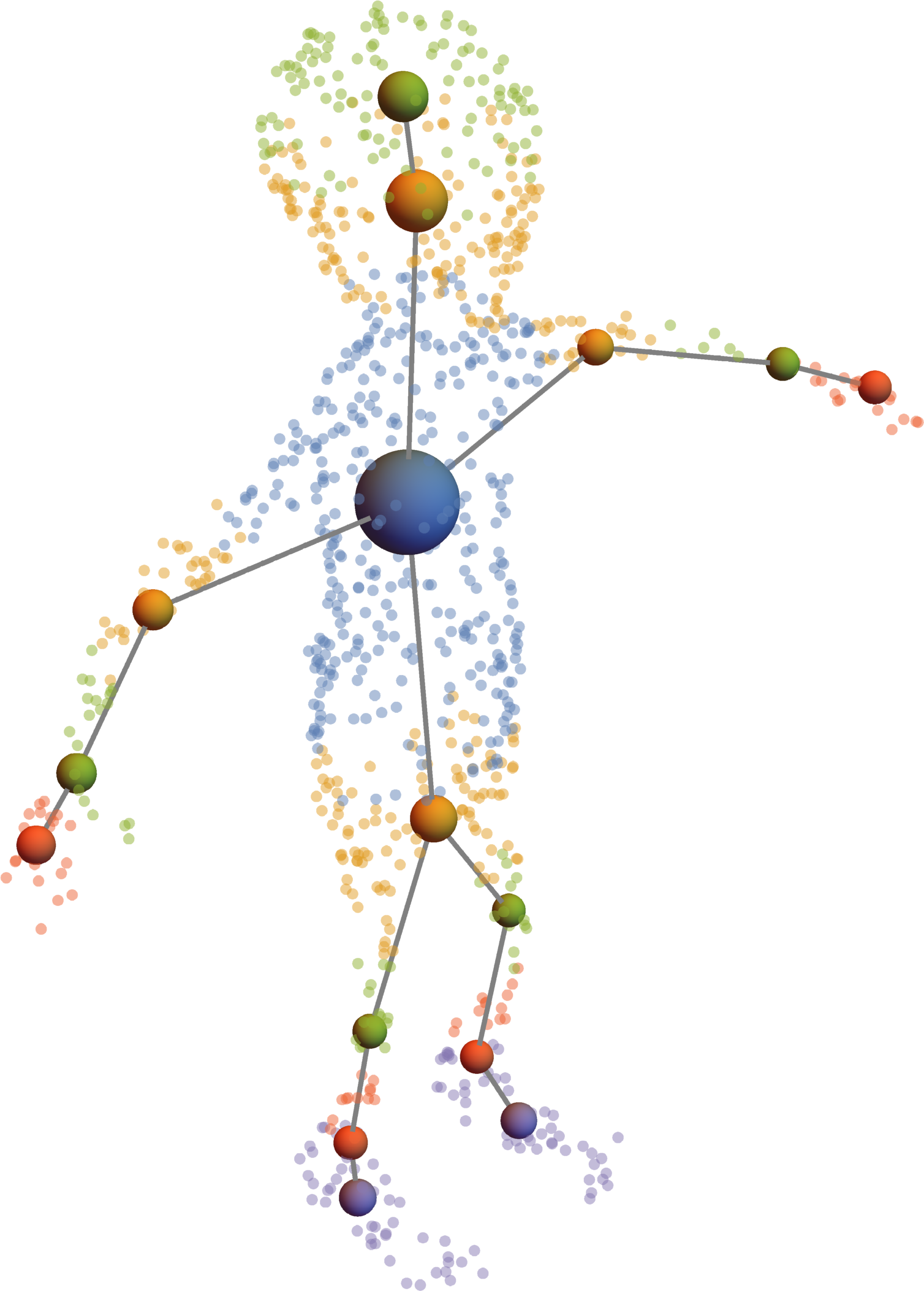}}} \\
  \subfloat{{\includegraphics[width=9cm]{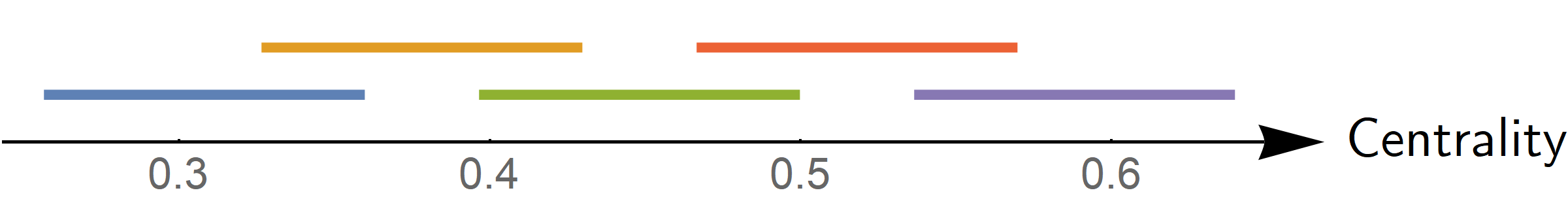}}}
  \caption{Result of Mapper algorithm applied to 1000 uniformly sampled points from the alien model (left),
   using centrality function as a filter, normalised by the number of data points: $f(x) =\frac{1}{1000}\sum_{y\in \text{data}}d(x,y)$.
  The resulting graph is overlayed onto the data (right), coloured by
    the corresponding intervals in the cover (bottom).}
  \label{alien2}
\end{figure}

\noindent
To obtain a more detailed representation of the alien with Mapper, we can refine
the resolution of the filter by, for example,  reducing the length of the intervals
in the cover. Furthermore, we can capture the shape of the data better through
use of another filter function, such as centrality. The resulting graph in
Figure \ref{alien2} still has five appendages, but has more vertices. \\

\begin{figure}[ht!]
  \centering
  \subfloat{{\includegraphics[width=5cm]{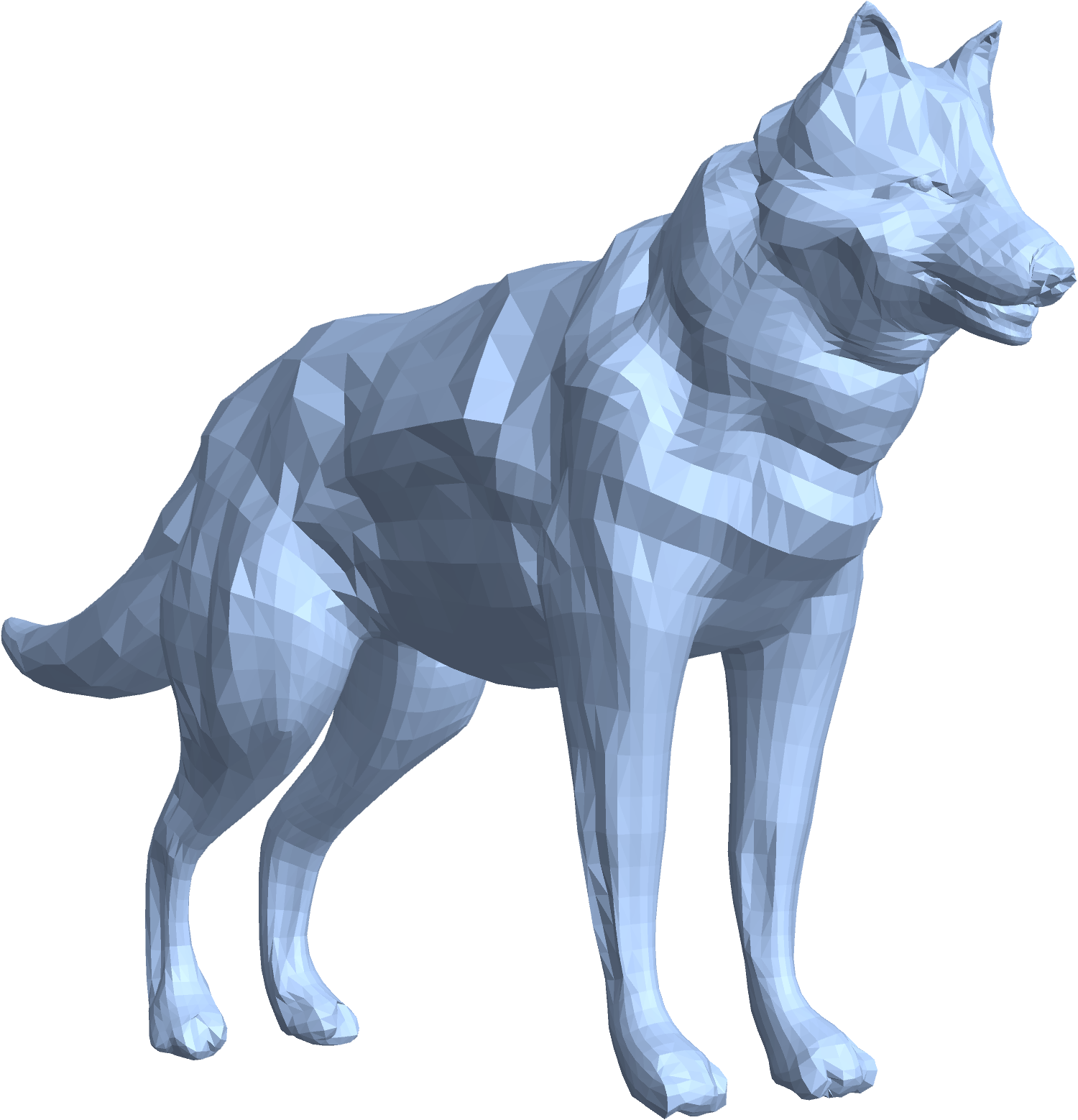}}}
  \quad
  \subfloat{{\includegraphics[width=5cm]{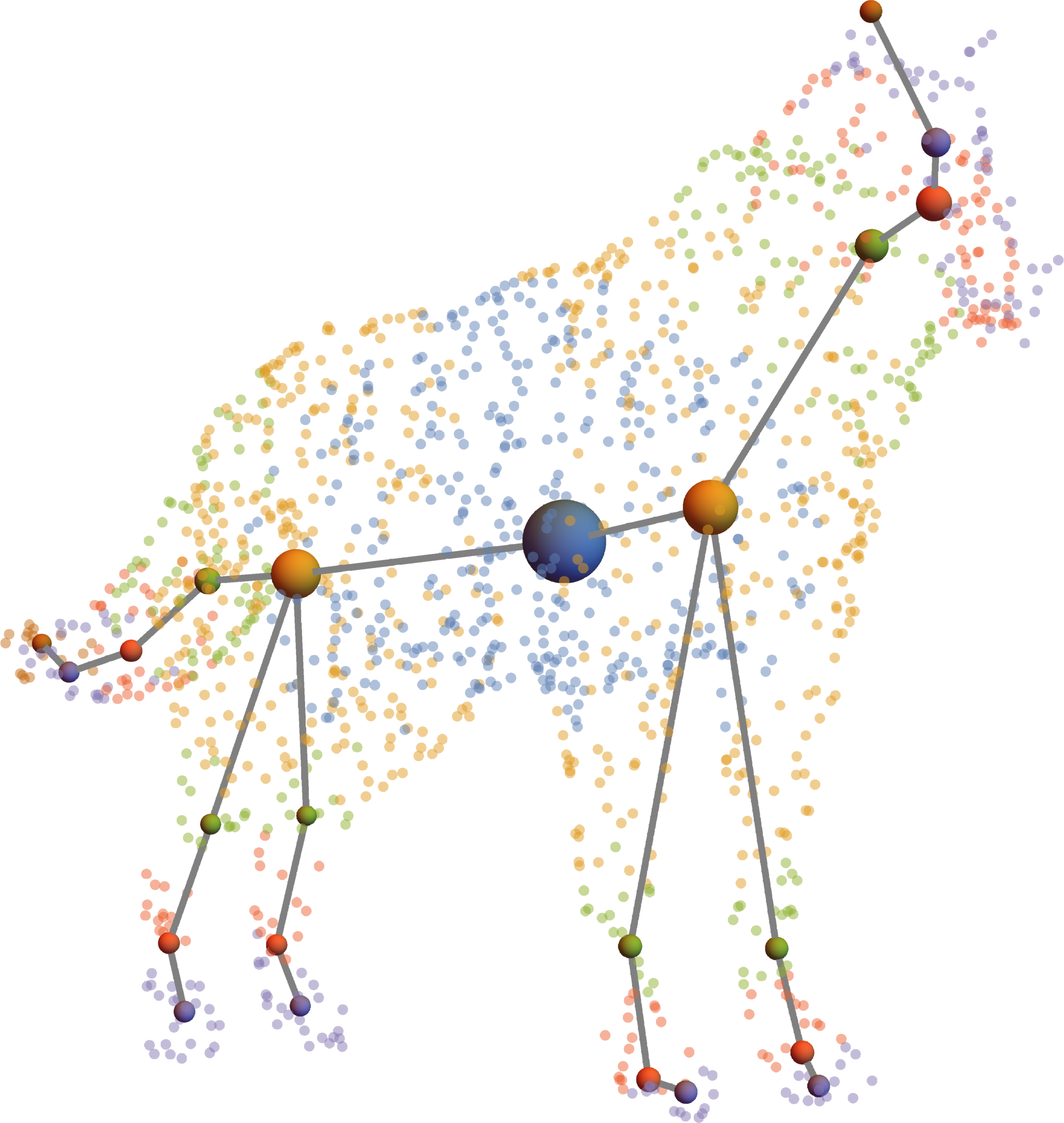}}} \\
  \subfloat{{\includegraphics[width=9cm]{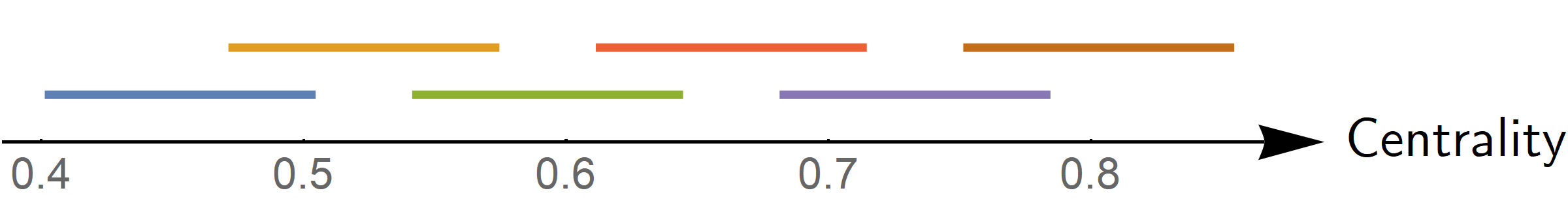}}}
  \caption{Result of Mapper algorithm applied to 1000 uniformly sampled points
    from the wolf model \cite{mnphmnmn_free_nodate} (left),
    using the centrality function as a
    filter. The resulting graph is overlayed onto the data (right), coloured by
    the corresponding intervals in the cover (bottom).}
  \label{wolf1}
\end{figure}

\noindent
Notice that despite using a different type of filter in Figure \ref{alien2}, we still obtain the same core structure of the alien, as we did with the height function.  In Figure \ref{wolf1}, we apply the algorithm to another 3D image, this time a wolf. Again, we see that Mapper identifies the basic form of the wolf, using the centrality filter.

\subsubsection{Illustrations of Mapper's robustness to noise}

The Mapper algorithm, like persistent homology, is perfectly capable of identifying
topological features of noisy datasets. Furthermore, it is now possible to
construct confidence regions for these topological features \cite{carriere_statistical_2017}. In this section, we
will briefly consider the effect of noise on the Mapper algorithm's output, to
motivate why it is important to understand Mapper's statistical properties.\\

\begin{figure}[ht!]
  \centering
  \subfloat{{\includegraphics[width=4cm]{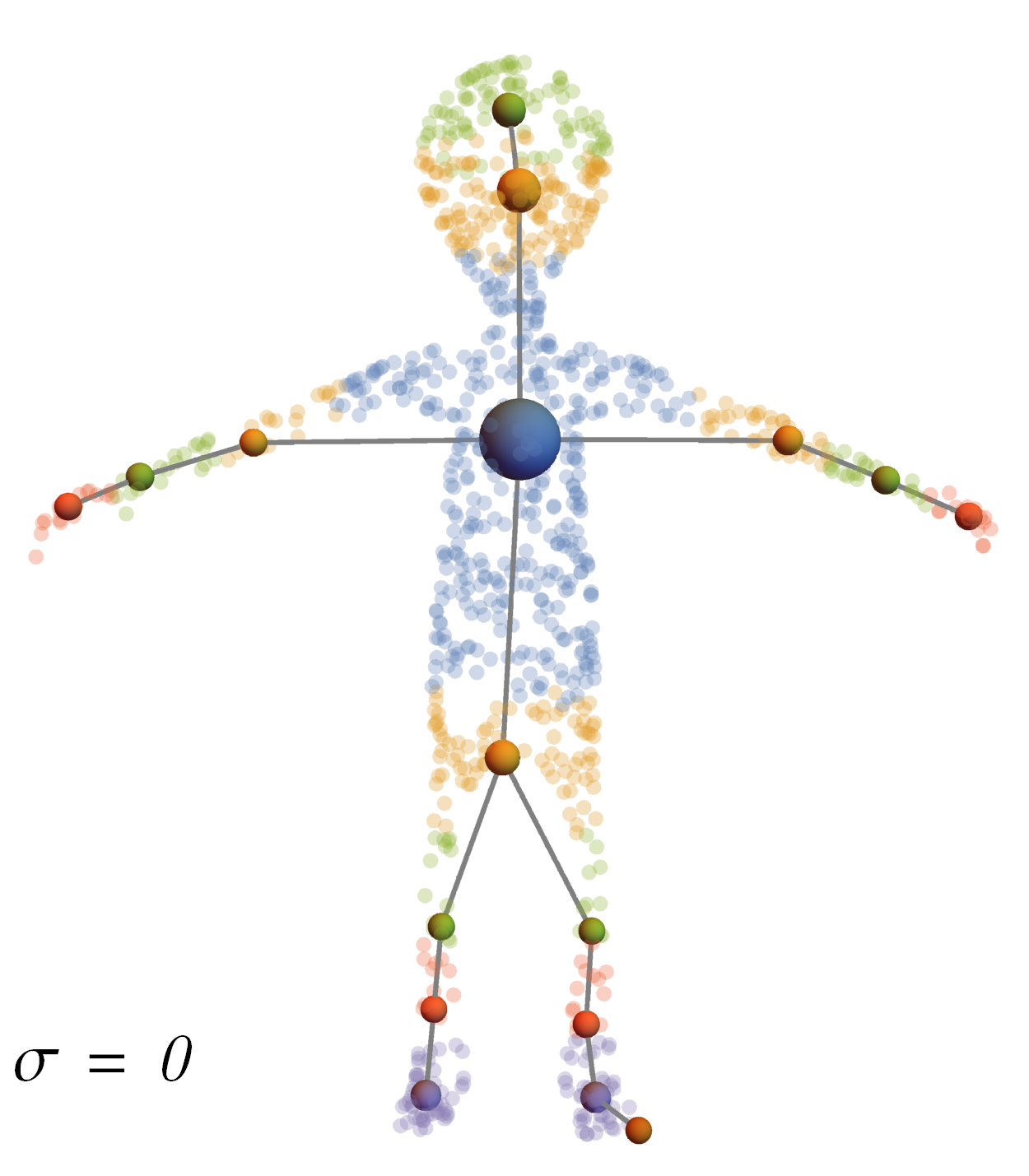}}}
  \quad
  \subfloat{{\includegraphics[width=4cm]{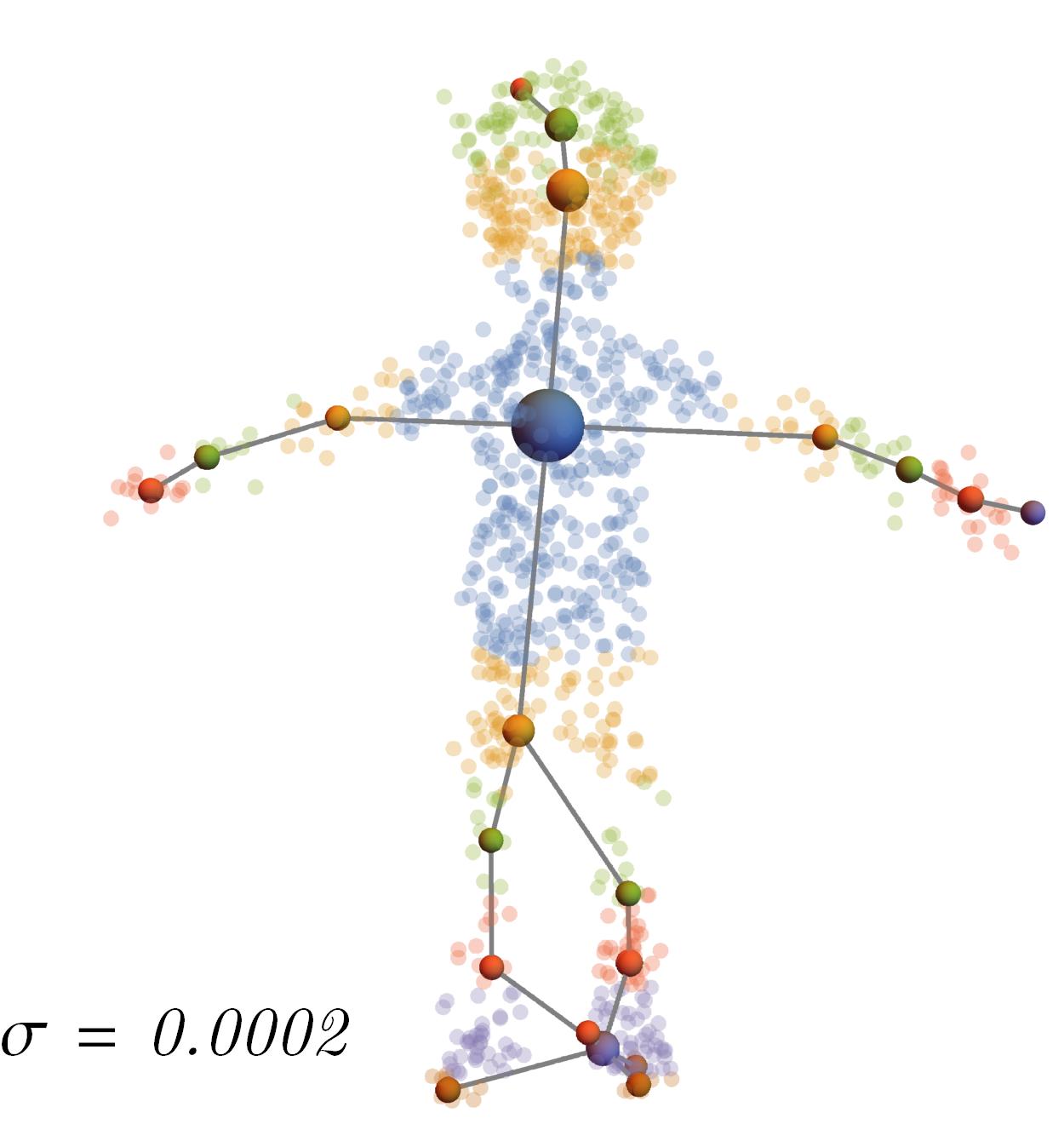}}} \\

  \subfloat{{\includegraphics[width=4cm]{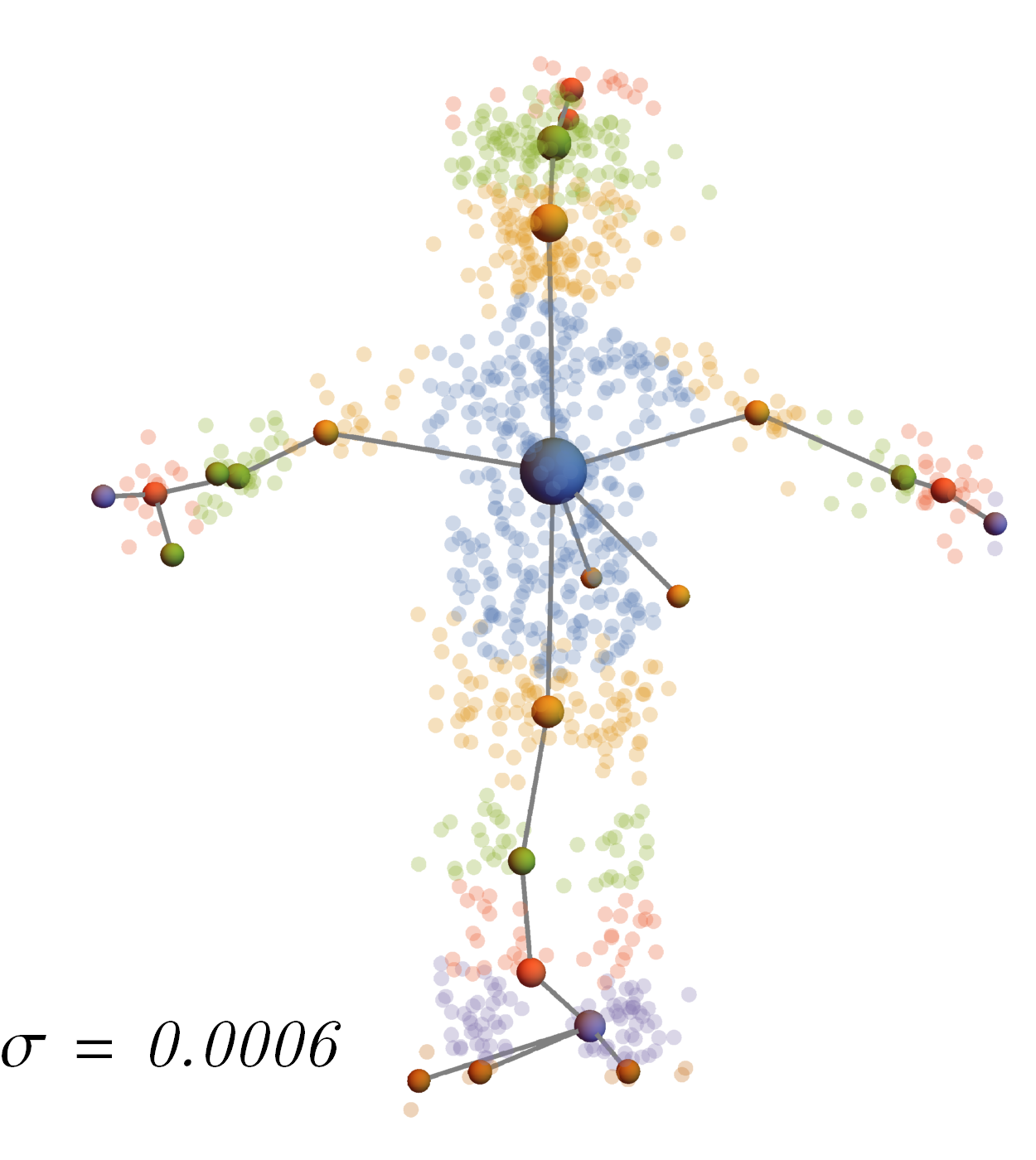}}}
  \quad
  \subfloat{{\includegraphics[width=4cm]{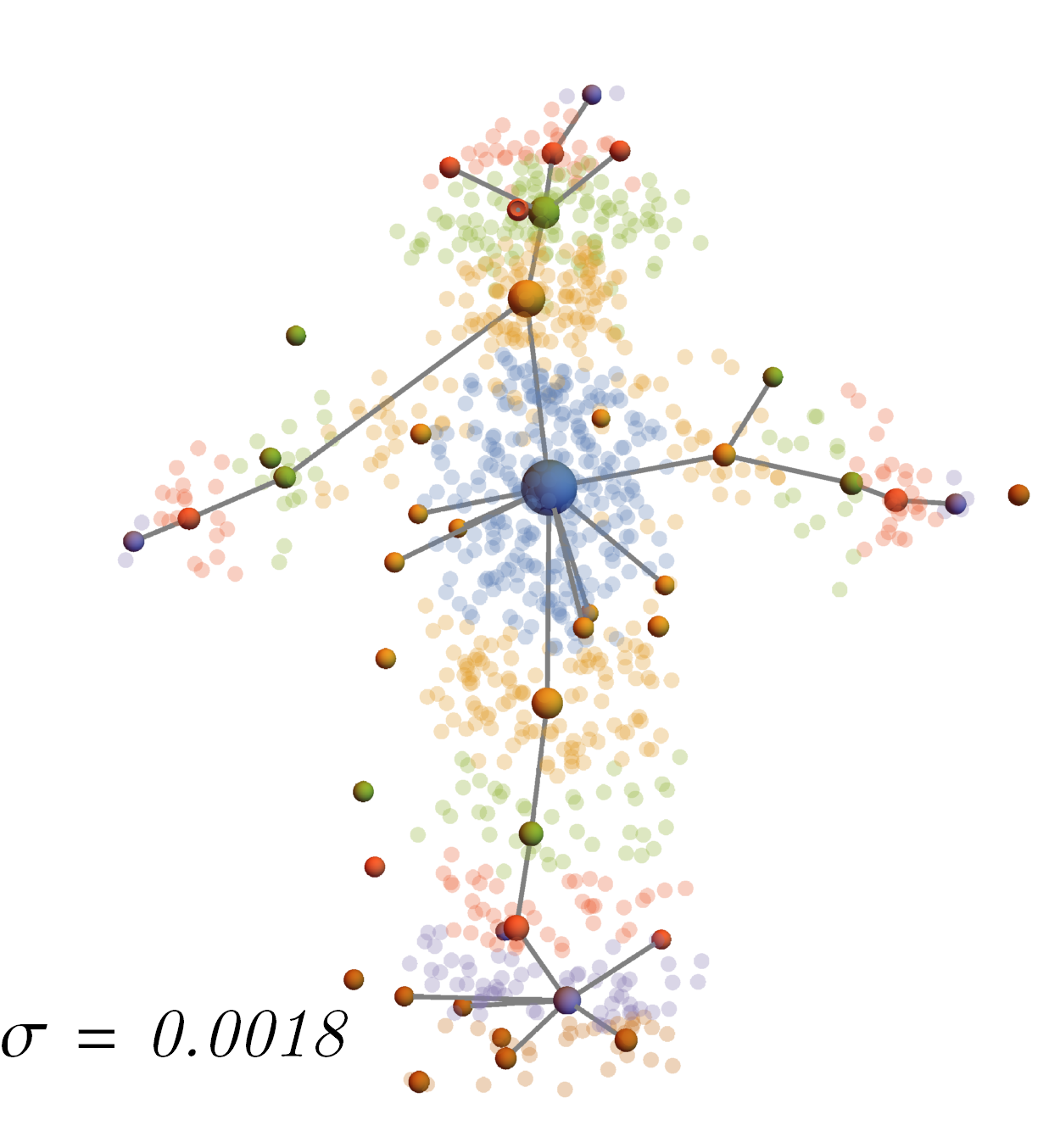}}} \\
  \subfloat{{\includegraphics[width=8cm]{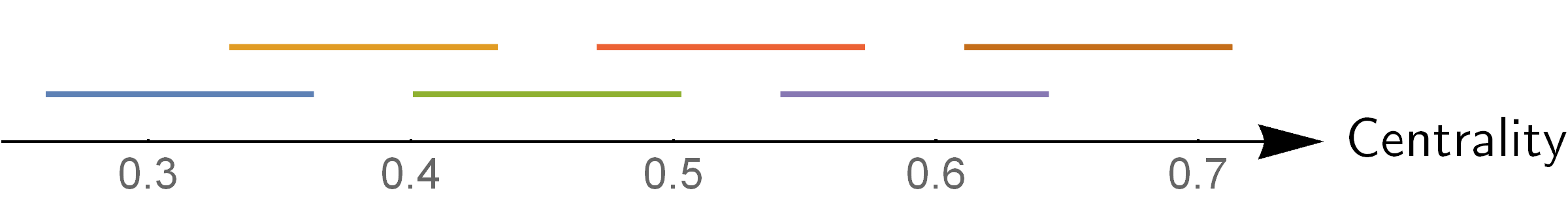}}}
  \caption{Mapper algorithm output for a vector of 1000 points, sampled uniformly
    from the alien object, using the normalised centrality filter function. A vector of
    1000 samples from a 3D Gaussian distribution was then added to the data
    vector, using variance $\sigma^2 I$ as indicated in each plot.}
  \label{aliennoise}
\end{figure}

\noindent
Returning to our alien example in Figure \ref{aliennoise}, let's now add some Gaussian
noise of increasing variance to study how the Mapper output changes. The
legs of the alien are not distinguished by the algorithm in the presence of
slight noise. This can primarily be attributed to the clustering method used, and
a consequence of the fact that the legs are connected in the $\epsilon$-neighbouring graph of the data used
for clustering (not pictured), which results in the clusters of points from each
leg being merged into one.\\

\begin{figure}[ht!]
  \centering
  \subfloat{{\includegraphics[width=4cm]{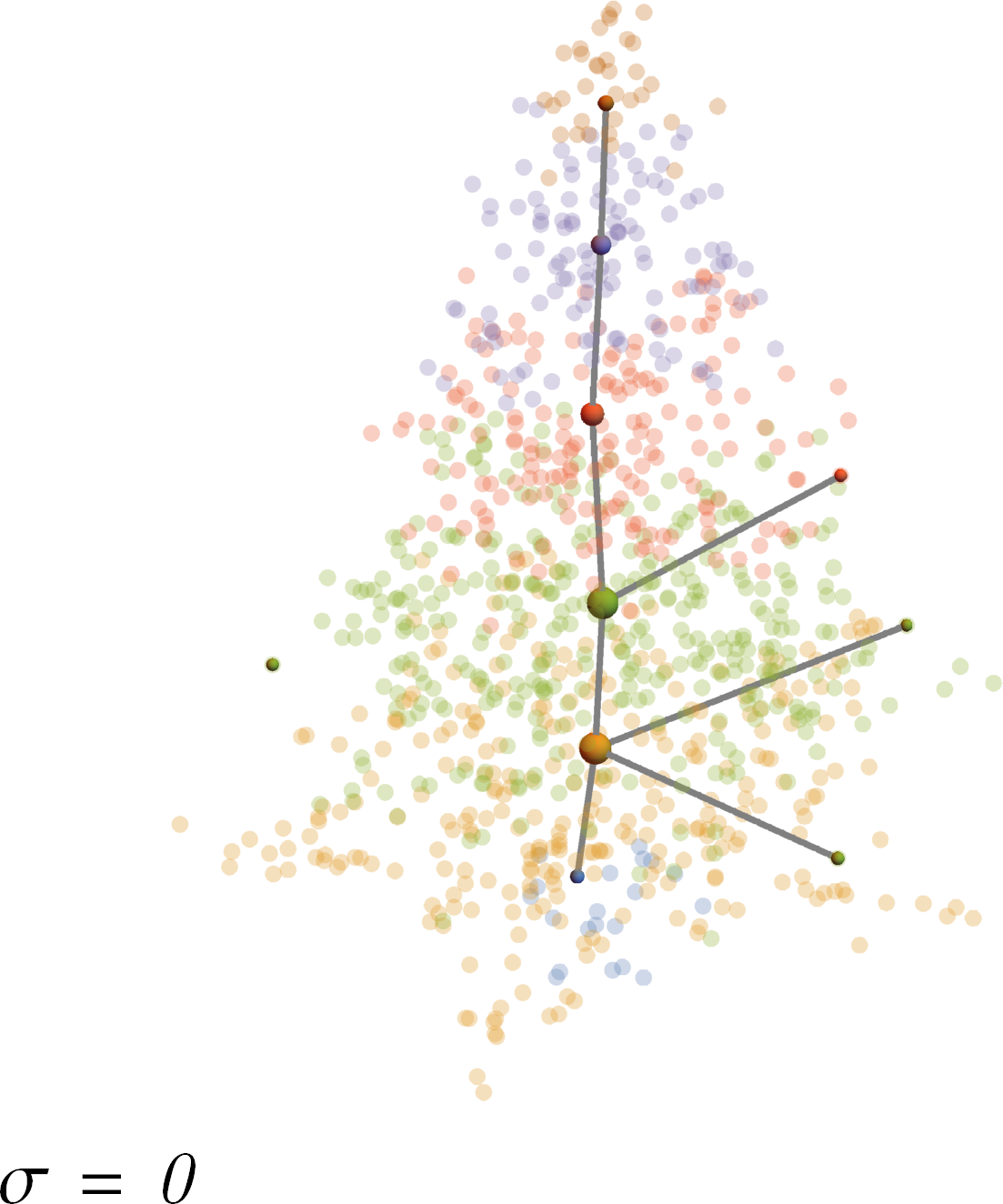}}}
  \quad                                    
  \subfloat{{\includegraphics[width=4cm]{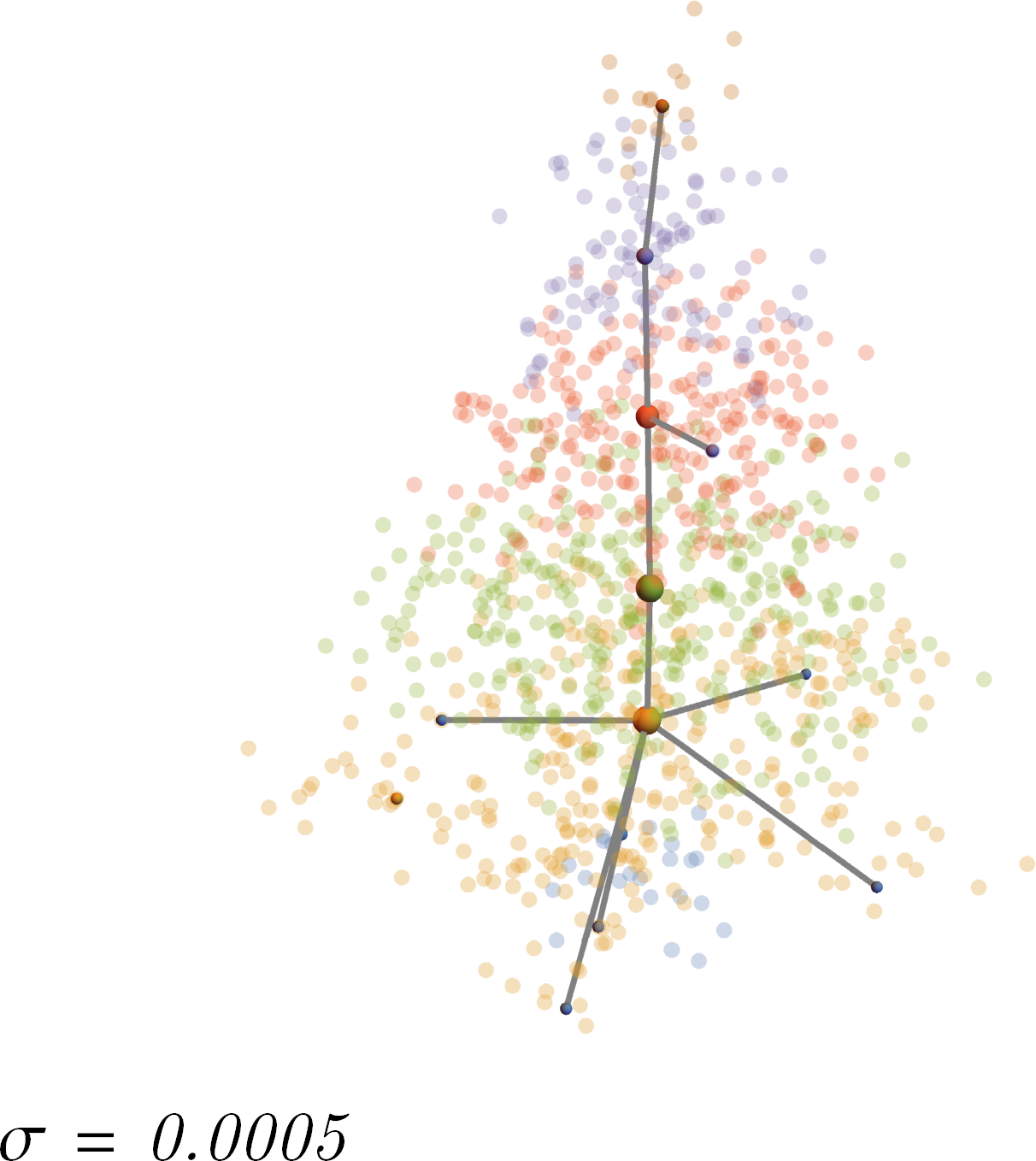}}} \\
                                          
  \subfloat{{\includegraphics[width=4cm]{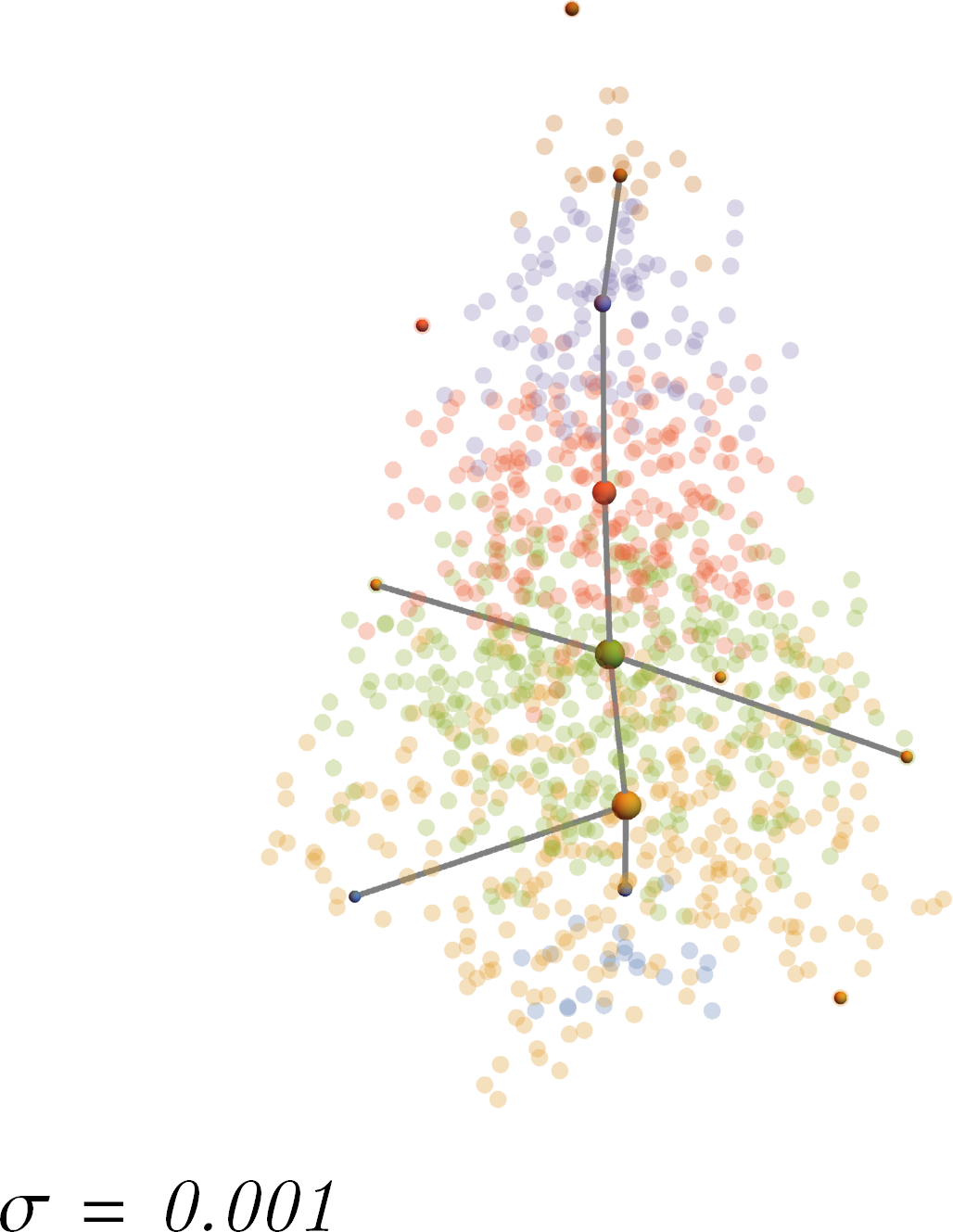}}}
  \quad                                  
  \subfloat{{\includegraphics[width=4cm]{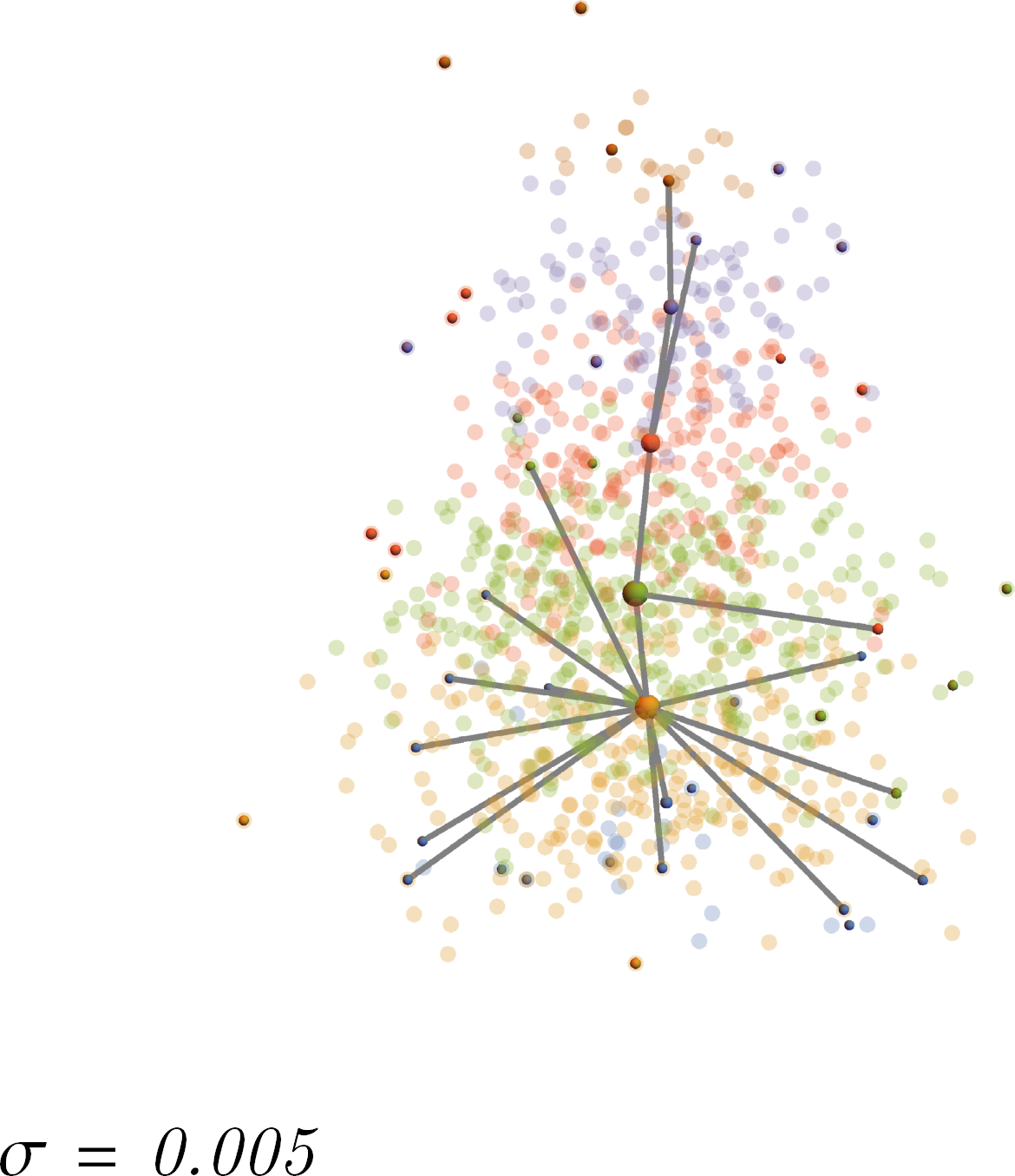}}} \\
  \subfloat{{\includegraphics[width=8cm]{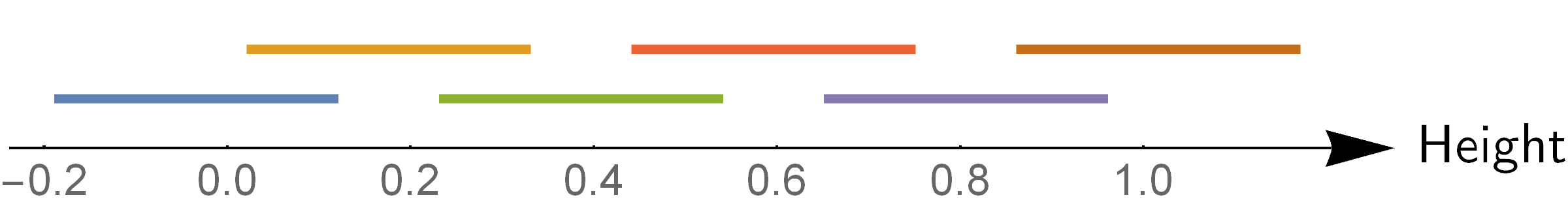}}}
  \caption{Mapper algorithm output for a vector of 1000 points, sampled uniformly
    from Christmas tree object\cite{printable_models_christmas_nodate}, using the normalised centrality filter function. A vector of
    1000 samples from a 3D Gaussian distribution was then added to the data
    vector, using variance $\sigma^2 I$ as indicated in each plot.}
 \label{christmas}
\end{figure}

\noindent
To check some of these conclusions, we tested Mapper on yet another 3D image; this time a Christmas tree. Here also it has problems identifying the branch structure of points sampled from the tree. Specifically, the outliers in Figure \ref{christmas} cause a proliferation of disconnected vertices in the output graph. Either the parameters need tuning, or outlier analysis needs to performed on the data before Mapper is applied. \\

\begin{figure}[ht!]
  \centering \includegraphics[width=\textwidth]{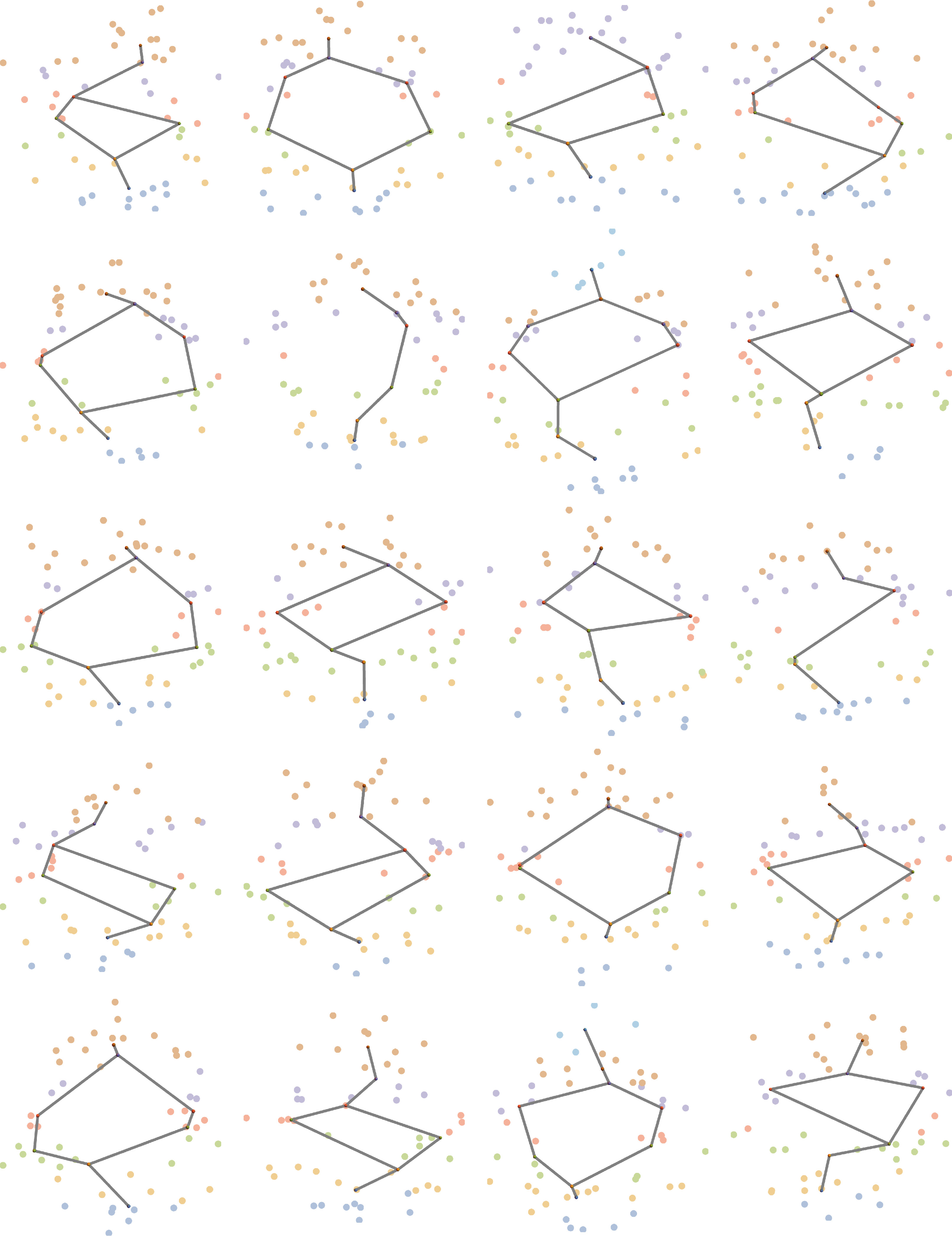}
  \caption{Mapper output for 24 samples of size 50 from a unit circle, with
    Gaussian noise of variance $\sigma^2 = 0.04$ applied. The cover had
    intervals of length 0.8, overlapping 50\%.}
  \label{loops1}
\end{figure}

\noindent
Notice that for the highest variance in Figure \ref{aliennoise}, many disconnected vertices are appear in the graph making it increasingly difficult for the algorithm to identify topological properties of the alien shape. However, the clustering method and other parameters are easily tweaked to circumvent this problem. Turning now to a simpler example, we consider points sampled from unit circle with Gaussian noise applied, of fixed variance. In Figure \ref{loops1}, we see that Mapper has correctly identified that there is a loop structure in the data in all but two of the 24 samples. The parameter choice here was critical; the filter requires a high gain (50\% overlap of intervals) in order to prevent disconnected vertices in the output. Alternatively, we could have applied persistent homology to these data to identify the loops.\\

\noindent
To summarize, the Mapper output can be thought of as an estimator of the topology of
the dataspace with respect to a filter function. From these examples, we have seen that Mapper may fail to identify the topology of the shape from which data are sampled, especially in the presence of outliers and noisy data. Hence, Mapper's estimate is not perfectly precise. As a result, confidence regions for the estimates are critical if one is to base ones conclusions on Mapper output. 

\clearpage

\section{TDA and FRBs}

As a final illustration of the methods in this introductory article, we turn our attention skywards. From exoplanets, to black holes [{\bf M87}], astrophysical phenomena provide some of the most fertile grounds for ``big data". Traditionally the exclusive domain of astrophysicists, the sheer volume of data captured by telescopes like the Square Kilometer Array (SKA) has meant that analysis of such large data sets is seeing greater input from data scientists that bring sophisticated new techniques from statistics and machine learning to the table. We believe strongly that TDA is one such powerful tool to add to the arsenal of astrophysicists. Toward the goal of making this case, we end this article with a foray into one such astrophysical phenomenon, the so-called fast radio bursts (FRBs) that are as remarkable as they are mysterious. More so than the rest of the paper, we consider this section an exercise in {\it experimental mathematics} - let's use the powerful tool that is TDA to see what patterns we can uncover in the existing FRB data. Our choice of FRBs over any other of the multitude different data sets is two-fold; (i) the field is very much in its infancy with less than a hundred catalogued events\footnote{Although we are assured that this number is set to increase dramatically in the next few months.}, so we expect that the likelihood of TDA noticing {\it new} patterns in the data is higher, and (ii) with a sparse, highly parameterised dataset, FRBs are more susceptible to analysis with limited computing resources.

\subsection{A lightning introduction to FRBs}

Fast radio bursts are short (mostly less than 5 ms in duration) and powerful (0.1 - 100 Jy in spectral flux density) radio pulses \cite{popovFastRadioBursts2018}. The exact nature of their source is a mystery. Proposed candidates include emissions from magnetars, to superconducting superstrings, to radio pulsars\cite{Metzger:2019una, Costa:2018zwb, Lieu:2016hfw}. At the time of writing however, only 52 FRBs have been confirmed since their first discovery in 2007 \cite{petroffFRBCATFastRadio2016}. However, these are not thought to be rare events, and it is estimated that thousands of FRBs are observable across the night sky every day \cite{popovFastRadioBursts2018}.\\

\noindent
The theory that FRBs are caused by catalysmic events was shaken by the observation of FRB 121102: the first known repeating FRB. Whether this implies multiple types of sources of FRB emissions is not yet clear, partly because FRB sources are only localised to within 3-10 arcminutes in the night sky, making the identification of FRBs within a particular galaxy very difficult \cite{calebOneSeveralPopulations2018}. Nevertheless, astronomers are abuzz with excitement about the observational potential of this new class of object. Various observed, estimated attributes and viable theoretical models of FRBs have been meticulously documented and are available in FRB catalogues \cite{petroffFRBCATFastRadio2016, Platts:2018hiy}.\\

\noindent
With the goal of identifying patterns in the FRB data using TDA, we need to choose a subset of these attributes which best characterise the observations.These attributes will parameterize the FRB data space. One such attribute is of course the location of the FRB. In the absence of precision redshift measurements, the center of the telescope beam in the night sky at the time of observation, in galactic coordinates, is usually taken as an approximation to the location of the FRB source. All telescopes record three basic attributes of the FRB signal \cite{petroffFRBCATFastRadio2016}:
\begin{itemize}
\item \textbf{Dispersion Measure (DM)}: Essentially, as a result of its propagation
  through dilute plasma, the energy in an FRB is staggered in its arrival at the
  observer, with higher frequencies arriving first. The dispersion measure is the
  integrated electron column density between the telescope (O) and the burst (S)
  \cite{petroffFRBCATFastRadio2016}. For
  example, suppose that we find ourselves $d$ parsecs away from the emitter of the FRB;
  consider a column extending from us to the FRB source with cross sectional area of 1
  $cm^2$. If the average density of free electrons in the column is $\langle n_e\rangle$
  electrons per $cm^3$, then
  \begin{equation*}
     DM = \int_{\mathrm{S}}^{\mathrm{O}}n_e\, dl = \langle n_e\rangle\cdot d,
  \end{equation*}
  i.e. the DM is proportional to the distance between us and the FRB source and can serve as a proxy therof.

\item \textbf{De-dispersed signal width}: The observed, de-dispersed width of the
  FRB pulse in ms. De-dispersion is carried out using the DM of the signal. 

\item \textbf{Signal to noise ratio}: 
  In radioastronomy, this is expressed as the
  ratio
  \begin{equation*}
    \text{SNR} = \frac{T_A}{\sigma_{\bar{T}}},
  \end{equation*}
  where $T_A$ is the system temperature contributed by the radio signal source
  (i.e. the `signal') and $\sigma_{\bar{T}}$ is the uncertainty in the measured
  mean temperature (i.e. the `noise'), which is proportional to the system
  temperature $T_{sys}$ \cite{simonettiRadioAstronomyFundamentals2010}.
\end{itemize} 
Further characterisations of an FRB, such as the linear and circular polarisation fractions, are only available for five FRBs in the FRB catalogue at time of writing and therefore do not provide a reliable parameterization over the full data set. Similarly, reliable measurements of flux density are not yet available, although an estimate thereof can be made from other observed parameters \cite{petroffFRBCATFastRadio2016}.

\subsection{Applying TDA to FRB data}
We will now use the FRB catalogue to illustrate how TDA tools can visualise data to reveal patterns. We will start with the most obvious: the identification of possible patterns in the distribution of FRBs in the night sky. 

\subsubsection{Spatial patterns in the FRB data}
\begin{figure}[ht!]
  \centering \includegraphics[height=12cm,width=10cm]{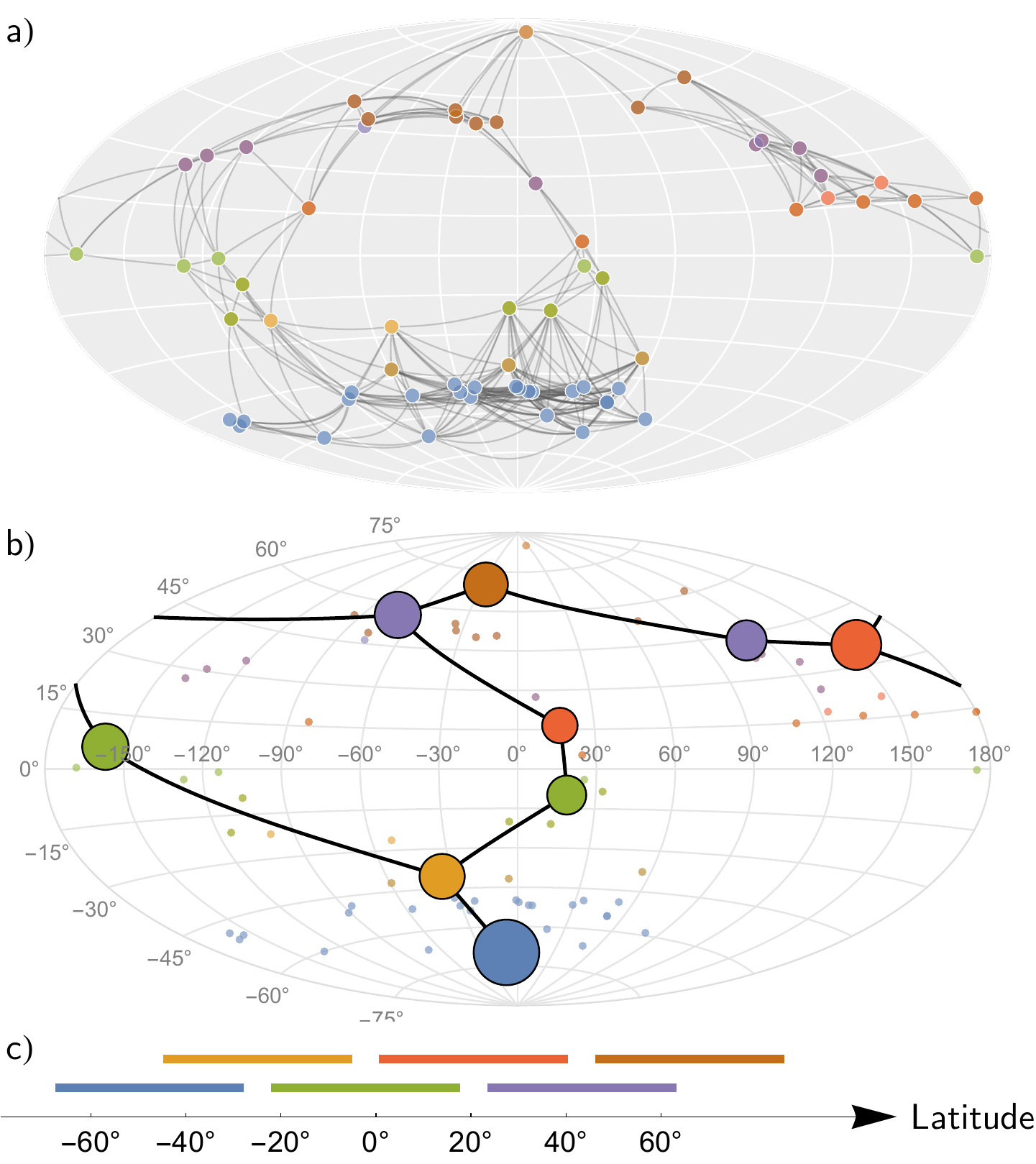}
  \caption{The nearest neighbour graph depicted in a), with $\epsilon = 45^\circ$, was used
    in the clustering method to obtain the Mapper graph of b). The great circle distance was
    chosen as the dissimilarity measure. The data were filtered by latitude, using a cover
    of five intervals with overlapping percentage 40\% (c). The data points in a) and b), as
    well as the vertices of the Mapper graph, are coloured by the interval(s) in c) which
    they correspond to.}
  \label{mapper_latitude}
\end{figure}

Our first task is to choose a metric for the spatial FRB data, whose positional data is given in longitude and latitude on the galactic sphere. A natural choice is the great-circle distance, which is the shortest distance between any two points on a sphere. To apply the mapper algorithm, we will use the clustering method based on a nearest-neighour graph, as in our previous examples. Figure \ref{mapper_latitude}a) depicts this graph. Figure \ref{mapper_latitude}b) gives the result of the mapper algorithm layed over the data, and filtered by latitude. Note that the edges of the graph are paths of constant bearing on the sphere. This is why they are curved when projected onto the plane.\\

\begin{figure}[ht!]
  \centering \includegraphics[height=5cm,width=8cm]{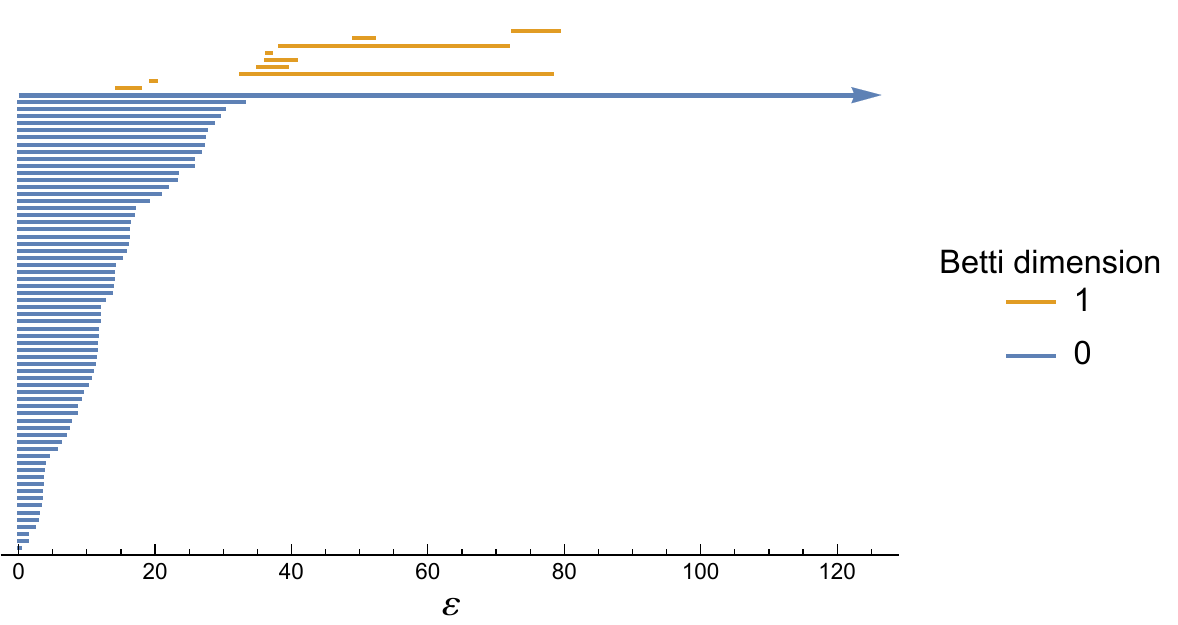}
  \caption{Barcode diagram of the spatial FRB data, using the great-circle distance as the
    metric. Orange lines indicate 1-holes.}
  \label{frb_barcode}
\end{figure}

\noindent
It is evident from Figure \ref{mapper_latitude} that the FRBs encircle two large areas of the celestial sphere: the north pole and much of the western hemisphere. As we have seen, persistent homology is an effective method to identify such holes in the data. The output of our \texttt{Mathematica} barcode generator is plotted in Figure \ref{frb_barcode}, and a quick glance at it corroborates this conclusion. There are indeed two 1-dimensional holes which persist over a large range 
        of $\epsilon$. \\
        
\noindent
How consequential is this ``pattern" in the FRB data set? Sadly, not very. All observational intuition seems to suggest that FRB observations in the very near future will eventually fill out a uniform, or near uniform, distribution on the celestial sphere. As such, we should expect that these ``circles in the sky" will not persist with a more dense data set.\\ 
        
\begin{figure}[ht!]
  \centering \includegraphics[height=5cm,width=8cm]{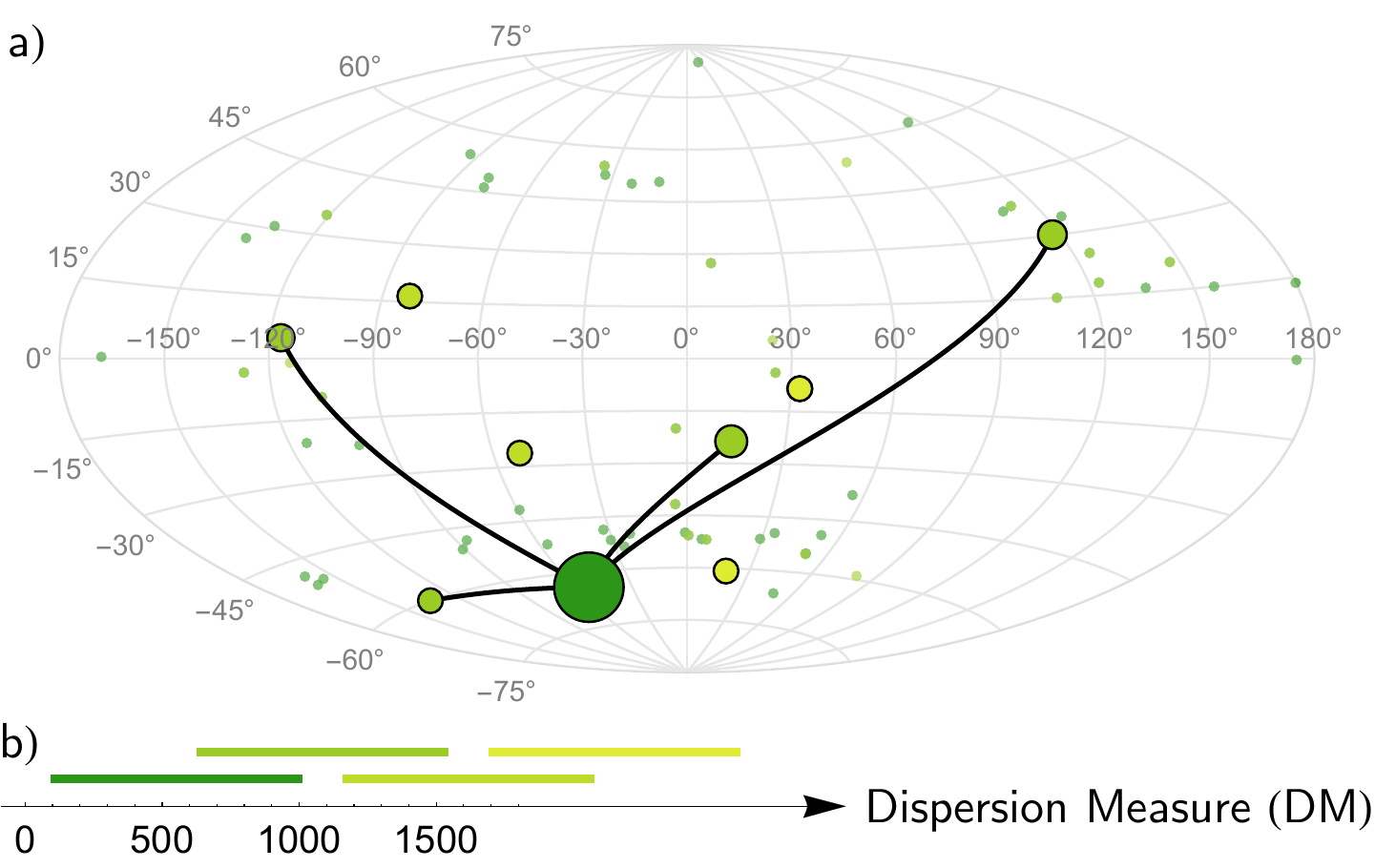}
  \caption{The Dispersion Measure (DM) of the FRBs was used as the filter for
    the Mapper algorithm, clustering using the same nearest-neighbour graph as
    in Figure \ref{mapper_latitude}a). The resulting Mapper output in a) is
    again coloured by the interval colours in b). Cover intervals were chosen
    with overlapping percentage of 40\%.}
  \label{mapper_DM}
\end{figure}

\begin{figure}[ht!]
  \centering \includegraphics[height=5cm,width=8cm]{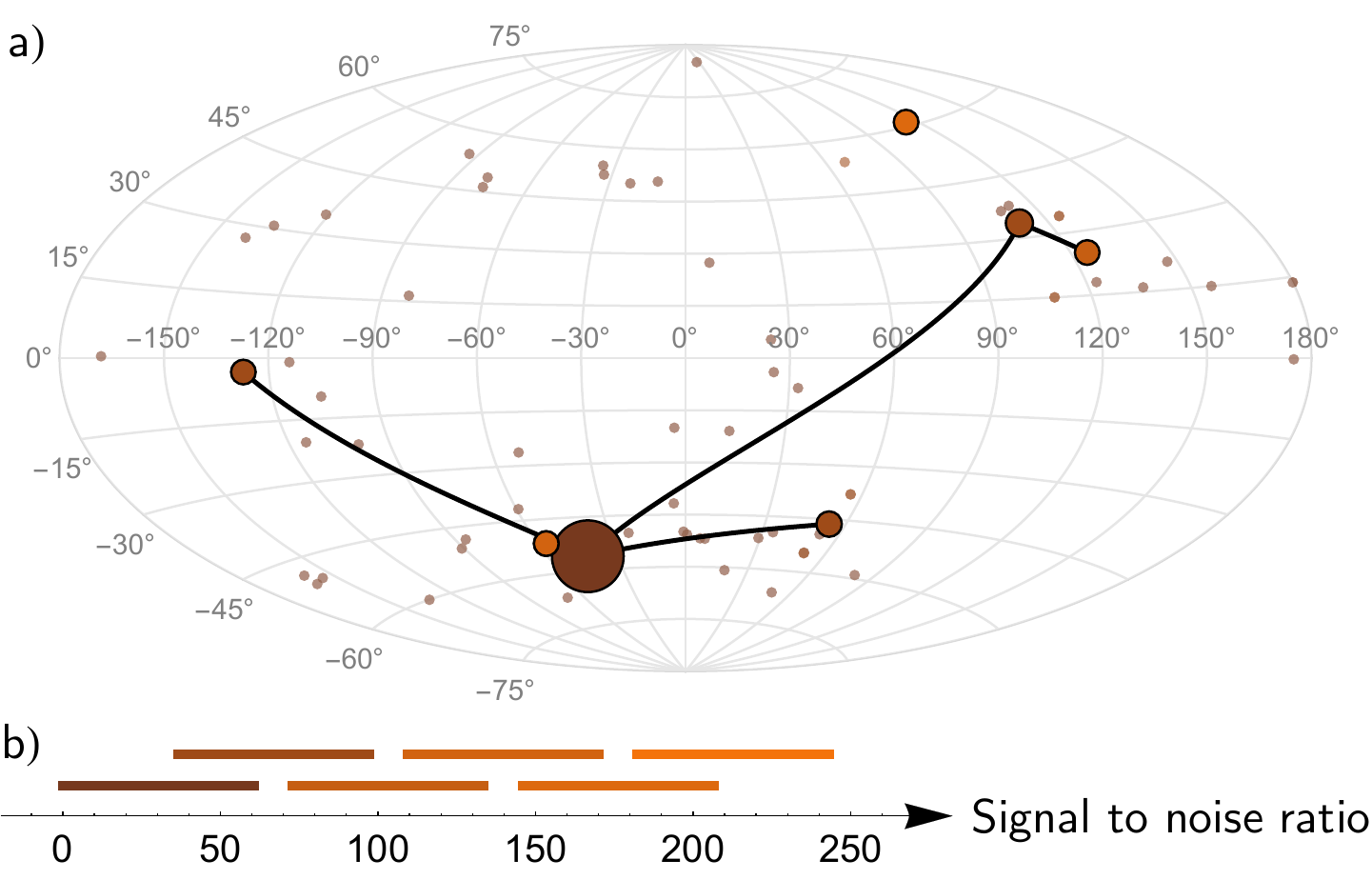}
  \caption{The FRB data was filtered by signal to noise ratio and clustered using the same nearest-neighbour graph as
    in Figure \ref{mapper_latitude}a). Data and vertices in the Mapper output in
  b) are again coloured by their corresponding interval(s) in b). Cover intervals were chosen
    with overlapping percentage of 40\%.}
  \label{mapper_SNR}
\end{figure}

\noindent        
We can go deeper into the structure of the FRB catalogue by feeding more variables into the Mapper algorithm. In particular, filtering by dispersion measure and signal-to-noise ratio produce the graphs in Figures \ref{mapper_DM} and \ref{mapper_SNR} respectively. Noteably, neither of these Mapper outputs display any interesting topological features with their nearest-neighbour graphs exhibiting only tree structures. This is a feature of a sparse data set and the fact that, even restricting ourselves to a  3-dimensional data space, each of these plots corresponds to just a single 2-dimensional slice of this 3-volume. Much more interesting would be the correlations between celestial position, dispersion and signal-to-noise ratio afforded by applying Mapper to the full data space. Motivated by the hope that TDA could well produce some organizational principle for the slew of forthcoming FRB data, we postpone this analysis for future work.

\section{Conclusion}

As a branch of applied mathematics, topological data analysis has already proven its utility many times over. Most of this has taken place in the exciting arena of systems biology where it has led to some truly remarkable developments in virology \cite{Chan18566} and oncology \cite{Nicolau7265}. Surprisingly, apart from some recent interesting work by Cole and Shiu \cite{Cole:2017kve, Cole:2018emh}, the uptake of TDA among the physics and astronomy communities has been less than whelming, especially since the value of topological data analysis lies in the fact that it is a specialised tool for the analysis of vast, high-dimensional datasets that are rapidly permeating all areas of physics and astronomy. This brief introduction is our modest attempt to ameliorate the situation by first reviewing details of the mathematical foundations of the field in a way we hope is accessible to a broad audience and then sketching some interesting applications that we have in mind for the near future. Briefly, some other interesting future problems suitable for topological data analysis are,
\begin{itemize}
   \item {\bf In astrophysics:} As we have discussed, FRBs are a whole new set of astrophysical phenomena that are only now beginning to make it onto the proverbial radar. It is not surprising then that the FRB dataset is sparsely populated, with few cataloged parameters and we have no doubt that with several large radio telescope missions on the horizon, this drought of data with soon be over. It would be of obvious interest to revisit the FRB dataset in the light of new discoveries by, for example the CHIME collaboration or, in the near future, the SKA. In the meantime, there are at least two more large, more well understood, datasets to consider: {\it exoplanets} and {\it gamma ray bursts} (GRBs). Uncovering structure in either of these datasets would be of great importance. 
   \item {\bf In condensed-matter physics:} Topological quantum materials are novel quantum states of matter that exhibit linear electronic response in the bulk and anomalous gapless states on their boundaries. These properties have led to tremendous excitement among materials scientists that topological quantum matter may finally provide a path for what lies beyond silicon valley. However while there has been enormous development in the theoretical understanding of these quantum systems, expeimental progress has been hampered by the difficulties in the practical computation of topological invariants for various compounds. This all changed in 2018 when, in a remarkble paper, Zhang {\it et.al.} revealed their \texttt{Materiae} catalogue \cite{2018arXiv180708756Z}, a searchable database of 39519 materials, more than 8000 of which are viable topological material (either topological semimetals, topological insulators or topological crystalline insulators). Like any database, \texttt{Materiae} begs out for one (or more) organizational principle\footnote{We have in mind here something akin to the periodic table of elements, but for topological quantum matter}. TDA, with its ability to distinguish patterns in high-dimensional datasets, may well be the tool for the task.
   \item {\bf In cosmology:} The first steps toward utilizing TDA to understand large scale structures in the cosmic microwave background (CMB) were recently taken by Cole and Schiu in \cite{Cole:2017kve} where they found that, using persistence diagrams applied to CMB temperature anisotropy data, they could constrain various cosmological parameters. At their most optimistic, they estimate that their persistence analysis should be able to constrain local non-Gaussianity in the CMB to $\Delta f_{NL} = 35.8$ at the 68\% confidence level. The article itself is more a proof-of-principle than detailed precision computation, so the authors rightly point out that these numbers should not be taken too seriously. However, their persistence computations reveal just how much can be accomplished using topological methods. Of particular interest to us would be to carry out a persistence or Mapper analysis on the topology of the large scale structure of spacetime itself.
\end{itemize}
This is not to say, however, that TDA is a universal panacea for data-intensive science. Indeed, the broad utilization of TDA methods come with their own challenges, two of which we now discuss.

\begin{itemize}
\item{\bf Computational challenges:} Persistent homology is computationally expensive.
  For example, to construct a barcode diagram for a set of just 20 points in
  $\mathbb{R}^2$ can require the construction of a simplicial complex of 10+ dimensions,
  with hundreds of simplices! Our own persistent homology code implemented in
  \texttt{Mathematica} struggled with simple data sets, while being incapable of
  handling more complex, real-world data. Hence, highly efficient implementations of the
  algorithms, written in a high performance language such as \texttt{C++}, are essential
  if persistent homology is to be applied to big datasets. On the other hand, the Mapper
  algorithm proved to be much easier to implement, and much faster to run, than
  persistent homology. Creating Mapper graphs for 2000 sample size datasets took only a
  few seconds on a mid-range laptop. Consequently, our \texttt{Mathematica}
  implementation of Mapper, is capable of handling moderately-sized real-world data.
  However, here too, high-dimensional datasets will likely require more efficient
  implementations.

\item{\bf Statistical challenges:} The basic implementations of TDA, as described in
  this article, are purely descriptive of the data; there is no statistical inference
  involved. However, both branches of TDA exhibit some innate robustness to noise.
  Indeed, we have seen that persistent homology and Mapper can both uncover the topology
  of a shape from which data was sampled, even if mild noise is added to the dataset.
  The problem, however, is that in real-world datasets, we generally do not know the
  topology of the space from which our data is sampled. This makes the checking of
  whether the output of TDA algorithms is representative of the dataspace topology very
  difficult. In statistics, given an estimator with some distribution, one can construct
  confidence regions for estimates of data parameters. The output of TDA algorithms can
  be thought of as estimators of the dataspace's topology, and hence it is desirable to
  construct confidence regions for these estimates too. Fortunately, methods for
  constructing these regions have recently been created for both persistent
  homology \cite{fasy_confidence_2014} and
  Mapper \cite{carriere_statistical_2017}, providing some much needed tools to
  conduct rigorous statistical inference on the topology of dataspace.
\end{itemize}
These caveats in place, TDA is a large powerful hammer and, if nothing else, we hope to have stimulated a search for more nails.
 
\section{Acknowledgements}
We would like to thank Jonathan Shock for collaboration on the early stages of this work and Bryan Gaensler and Amanda Weltman for their invaluable insights on FRBs.  JM is supported by the NRF of South Africa under grant CSUR 114599. DR acknowledges funding from the Harry Crossley Research Fellowship of the University of Cape Town. 


\end{document}